\documentclass{article}

\usepackage{csquotes}
\usepackage{fullpage, amsmath, amsthm, amssymb}
\usepackage{enumerate}
\usepackage{float}
\usepackage{color,  graphicx}

\newtheorem{definition}{Definition}
\newtheorem{theorem}{Theorem}
\newtheorem{corollary}{Corollary}
\newtheorem{lemma}{Lemma}

\newtheorem{claim}{Claim}
\newtheoremstyle{restate}{}{}{\itshape}{}{\bfseries}{~(restated).}{.5em}{\thmnote{#3}}
\theoremstyle{restate}
\newtheorem*{restate}{}

\newcommand{\algwidth}{0.97\textwidth}
\newcommand{\Fam}{\mathcal{F}}
\DeclareMathOperator{\mmod}{mod}
\renewcommand{\P}{\textbf{P}}
\newcommand{\Q}{\textbf{Q}}
\newcommand{\E}{\mathop{{}\mathbb{E}}}
\renewcommand{\S}{\textbf{S}}
\newcommand{\p}{\textbf{p}}
\renewcommand{\log}{\lg}

\newcommand{\R}{\mathbb{R}}
\newcommand{\eps}{\epsilon}
\newcommand{\path}{\mathsf{path}}
\newcommand{\suc}{\overrightarrow{\mathsf{adv}}}

\newcommand{\mnote}[1]{ \marginpar{\tiny\bf
            \begin{minipage}[t]{0.5in}
              \raggedright #1
           \end{minipage}}}
           
\newcommand{\GF}{GF}
\newcommand{\cA}{\mathcal{A}}
\newcommand{\cB}{\mathcal{B}}
\newcommand{\cC}{\mathcal{C}}
\newcommand{\cE}{\mathcal{E}}
\newcommand{\cI}{\mathcal{I}}
\newcommand{\cP}{\mathcal{P}}
\newcommand{\cG}{\mathcal{G}}
\newcommand{\cH}{\mathcal{H}}
\newcommand{\cM}{\mathcal{M}}
\newcommand{\cN}{\mathcal{N}}
\newcommand{\cX}{\mathcal{X}}
\newcommand{\cY}{\mathcal{Y}}
\newcommand{\cS}{\mathcal{S}}
\newcommand{\cQ}{\mathcal{Q}}
\newcommand{\cU}{\mathcal{U}}
\newcommand{\cT}{\mathcal{T}}
\newcommand{\cR}{\mathcal{R}}
\newcommand{\cL}{\mathcal{L}}
\newcommand{\cF}{\mathcal{F}}
\newcommand{\cZ}{\mathcal{Z}}
\newcommand{\cD}{\mathcal{D}}
\newcommand{\cJ}{\mathcal{J}}
\newcommand{\cK}{\mathcal{K}}
\newcommand{\cO}{\mathcal{O}}
\newcommand{\cW}{\mathcal{W}}

\newcommand{\bU}{\mathbf{U}}

\newcommand{\tS}{\tilde{S}}
\newcommand{\ORC}{\oplus\mathsf{ORC}}
\newcommand{\orc}{\mathsf{ORC}}
\newcommand{\poly}{\mathrm{poly}}
\newcommand{\ep}{\mathbf{U}}

\DeclareMathOperator{\supp}{supp}
\DeclareMathOperator{\rank}{rank}

\newcommand{\nz}{non-zero}
\newcommand{\AND}{\mathsf{AND}_k}

\def\mainfile{}

\begin{document}

\title{Crossing the Logarithmic Barrier for Dynamic Boolean 
 Data Structure Lower Bounds}

\author{
Kasper Green Larsen\thanks{Department of Computer Science, Aarhus
  University. Supported by MADALGO, grant DNRF84, a Villum Young
  Investigator Grant and an AUFF Starting Grant.} \and 
Omri Weinstein\thanks{Department of Computer Science, Columbia University.} \and 
Huacheng Yu\thanks{Department of Computer Science, Stanford University. Supported by NSF CCF-1212372.}}
\date{}

\pagebreak 
\maketitle 


\begin{abstract}
This paper proves the first super-logarithmic lower bounds on the
cell probe complexity of dynamic \emph{boolean} (a.k.a. decision) data structure problems, 
a long-standing milestone in data structure lower bounds.
 
We introduce a new method for proving dynamic cell probe lower bounds and use it to prove a $\tilde{\Omega}(\log^{1.5} n)$ lower bound on 
the operational time of a wide range of boolean data structure
problems, most notably, on the query time of dynamic range counting 
\emph{over $\mathbb{F}_2$} (\cite{Patrascu07}). Proving an $\omega(\lg n)$ lower bound for this problem was explicitly posed as one 
of five important open problems in the late Mihai P\v{a}tra\c{s}cu's
obituary~\cite{Thorup13}.  
This result also implies the first $\omega(\lg n)$
lower bound for the classical 2D range counting problem, one
of the most fundamental data structure problems in computational geometry and spatial databases.
We derive similar lower bounds for boolean versions of dynamic \emph{polynomial evaluation} and \emph{2D rectangle stabbing}, 
and for the (non-boolean) problems of \emph{range selection} and \emph{range median}. 

Our technical centerpiece is a 
new way of ``weakly" simulating dynamic  data structures using efficient \emph{one-way} communication protocols 
with small advantage over random guessing. 
This simulation involves a surprising excursion to 
low-degree (Chebychev) polynomials 
which may be of independent interest, and offers an entirely 
new algorithmic angle on the ``cell sampling" method of Panigrahy et al.~\cite{PTW10}. 
 
\end{abstract}

\ifx\mainfile\undefined
\documentclass{article}

\usepackage{csquotes}
\usepackage{fullpage, amsmath, amsthm, amssymb}
\usepackage{enumerate}
\usepackage{float}
\usepackage{color}

\newtheorem{definition}{Definition}
\newtheorem{theorem}{Theorem}
\newtheorem{corollary}{Corollary}
\newtheorem{lemma}{Lemma}
\newtheorem{proposition}{Proposition}
\newtheorem{claim}{Claim}
\newtheoremstyle{restate}{}{}{\itshape}{}{\bfseries}{~(restated).}{.5em}{\thmnote{#3}}
\theoremstyle{restate}
\newtheorem*{restate}{}

\newcommand{\algwidth}{0.97\textwidth}
\newcommand{\Fam}{\mathcal{F}}
\DeclareMathOperator{\mmod}{mod}
\renewcommand{\P}{\textbf{P}}
\newcommand{\Q}{\textbf{Q}}
\newcommand{\E}{\mathop{{}\mathbb{E}}}
\renewcommand{\S}{\textbf{S}}
\newcommand{\p}{\textbf{p}}
\renewcommand{\log}{\lg}

\newcommand{\R}{\mathbb{R}}
\newcommand{\eps}{\epsilon}
\newcommand{\path}{\mathsf{path}}
\newcommand{\suc}{\overrightarrow{\mathsf{adv}}}

\newcommand{\mnote}[1]{ \marginpar{\tiny\bf
            \begin{minipage}[t]{0.5in}
              \raggedright #1
           \end{minipage}}}
           
\newcommand{\GF}{GF}
\newcommand{\cA}{\mathcal{A}}
\newcommand{\cB}{\mathcal{B}}
\newcommand{\cC}{\mathcal{C}}
\newcommand{\cE}{\mathcal{E}}
\newcommand{\cI}{\mathcal{I}}
\newcommand{\cP}{\mathcal{P}}
\newcommand{\cG}{\mathcal{G}}
\newcommand{\cH}{\mathcal{H}}
\newcommand{\cM}{\mathcal{M}}
\newcommand{\cN}{\mathcal{N}}
\newcommand{\cX}{\mathcal{X}}
\newcommand{\cY}{\mathcal{Y}}
\newcommand{\cS}{\mathcal{S}}
\newcommand{\cQ}{\mathcal{Q}}
\newcommand{\cU}{\mathcal{U}}
\newcommand{\cT}{\mathcal{T}}
\newcommand{\cR}{\mathcal{R}}
\newcommand{\cL}{\mathcal{L}}
\newcommand{\cF}{\mathcal{F}}
\newcommand{\cZ}{\mathcal{Z}}
\newcommand{\cD}{\mathcal{D}}
\newcommand{\cJ}{\mathcal{J}}
\newcommand{\cK}{\mathcal{K}}
\newcommand{\cO}{\mathcal{O}}
\newcommand{\cW}{\mathcal{W}}

\newcommand{\bU}{\mathbf{U}}

\newcommand{\tS}{\tilde{S}}
\newcommand{\ORC}{\oplus\mathsf{ORC}}
\newcommand{\orc}{\mathsf{ORC}}
\newcommand{\poly}{\mathrm{poly}}
\newcommand{\ep}{\mathbf{U}}

\DeclareMathOperator{\supp}{supp}
\DeclareMathOperator{\rank}{rank}

\newcommand{\nz}{non-zero}
\newcommand{\AND}{\mathsf{AND}_k}

\begin{document}

\fi

\section{Introduction}

Proving unconditional lower bounds on the operational time of data structures in the cell probe model~\cite{Yao81} is one of 
the holy grails of complexity theory, primarily because lower bounds in this model are oblivious to implementation 
considerations, hence they apply essentially to any imaginable data structure (and in particular, to the ubiquitous word-RAM model). 
Unfortunately, this abstraction makes it notoriously difficult to obtain  
data structure lower bounds, and progress over the past three decades has been very slow.  
In the dynamic cell probe model, where a data structure needs to maintain a database under an ``online'' sequence of $n$ operations 
(updates and queries) by accessing as few memory cells as possible, a number of lower bound techniques have been developed. 
In~\cite{FS89}, Fredman and Saks proved $\Omega(\log n/\log\log n)$ lower bounds for a list of dynamic problems. 
About 15 years later, P\v{a}tra\c{s}cu and Demaine~\cite{PD04a,PD06} proved the first $\Omega(\log n)$ lower bound ever shown for an explicit 
dynamic problem.
The celebrated breakthrough work of Larsen~\cite{Larsen12a} brought a near quadratic improvement on the lower bound frontier, 
where he showed 
an $\Omega((\log n/\log\log n)^2)$ cell probe lower bound for the \emph{2D range sum} problem (a.k.a. weighted orthogonal range counting in 2D). 
This is the highest cell probe lower bound known to date.  

Larsen's result 
has one substantial caveat,
namely, it inherently requires the queries to have large ($\Theta(\log n)$-bit) output size.
Therefore, when measured per output-bit of a query, the highest lower bound remains only $\Omega(\log n)$ per bit (for dynamic connectivity due to 
P\v{a}tra\c{s}cu and Demaine~\cite{PD06}). 

In light of this,  
a concrete milestone that was identified 
en route to proving $\omega(\log^2n)$ dynamic cell probe lower bounds, 
was to \emph{prove an $\omega(\lg n)$ cell probe lower bound for \emph{boolean (a.k.a. decision)} data structure problems} (the problem was explicitly posed in~\cite{Larsen12a,Thorup13,LarsenThesis} and the caveat with previous techniques requiring large output has also been discussed in e.g.~\cite{Patrascu07,CGL15}). 
We stress that this challenge is \emph{provably} a prerequisite for going beyond the $\omega(\log^2n)$ barrier for general ($\Theta(\lg n)$-bit output) 
problems: Indeed, consider a dynamic data structure problem $\cP$ maintaining a database with updates $\cU$ and queries $\cQ$, where each 
query $q\in\cQ$ outputs $\log n$ bits. 
If one could prove an $\omega(\log^2 n)$ lower bound for $\cP$, this would directly translate into an $\omega(\log n)$ lower 
bound for the following induced dynamic \emph{boolean} problem $\cP^{\mathrm{bool}}$: $\cP^{\mathrm{bool}}$ has the same set of update operations 
$\cU$, and has queries $\cQ' := \cQ\times [\log n]$. 
Upon a query $(q, i)$, the data structure should output the $i$-th bit $(\cP(q,\cU))_i$ of the answer to the original query $q$ w.r.t the database $\cU$. 
An $\omega(\log n)$ lower bound then follows, simply because each query of $\cP$ can be simulated by $\Theta(\log n)$ queries of $\cP^{\mathrm{bool}}$, and the update time is preserved.
Thus, to break the $\log^2 n$-barrier for cell probe lower bounds, one must first prove a \emph{super-logarithmic} lower bound for some dynamic boolean problem. 
Of course, many classic data structure problems are naturally boolean (e.g., reachability, membership, etc.), hence studying 
decision data structure problems is interesting on its own. 

Technically speaking, the common reason why all previous techniques hitherto  
(e.g., \cite{Patrascu07,Larsen12a,weinstein:lbs}) fail to prove super-logarithmic 
lower bounds for dynamic boolean problems, is that they all heavily rely on each query revealing a large amount of information about the database. 
In contrast, for boolean problems, each query could reveal at most one bit of information,
and thus any such technique is doomed to fail. 
We elaborate on this excruciating obstacle and how we overcome it in the following subsection. 

In this paper, we develop a fundamentally new lower bound method and use it to prove the first super-logarithmic lower bounds for dynamic boolean data structure problems. 
Our results apply to natural boolean versions of several classic data structure problems. Most notably, 
we study a boolean variant of the dynamic \emph{2D range counting} problem. In 2D range counting, $n$ points are inserted one-by-one into an $[n] \times [n]$ integer grid, and given a query point $q = (x,y) \in [n] \times [n]$, the data structure must return the number of points $p$ \emph{dominated} by $q$ (i.e., $p.x \leq x$ and $p.y\leq y$). This is one of the most fundamental data structure problems in computational geometry and spatial database theory 
(see e.g., \cite{AgarwalSurveyRangeCounting04} and references therein). It is 
known that a variant of dynamic ``range trees" solve this problem using $O((\lg n/\lg \lg n)^2)$ amortized update time and $O((\lg n/\lg \lg n)^2)$ worst case query time (\cite{brodal:rangemedian}).  
We prove an $\tilde{\Omega}(\lg^{1.5} n)$ lower bound even for a boolean version, called \emph{2D range parity}, where one needs only to return the parity of the number of points dominated by $q$. This is, in particular, the first $\omega(\lg n)$ lower bound for the (classical) 2D range counting problem. 
We are also pleased to report that this is the first progress made on the 5 important open problems posed in Mihai P\v{a}tra\c{s}cu's obituary~\cite{Thorup13}.

In addition to the new results for 2D range parity, we also prove the first $\omega(\lg n)$ lower bounds for the classic (non-boolean) problems of dynamic \emph{range selection} and \emph{range median}, as well as an $\omega(\lg n)$ lower bound for a boolean version of \emph{polynomial evaluation}. We formally state these problems, our new lower bounds, 
and a discussion of previous state-of-the-art bounds in Section~\ref{sec:concreteproblems}. 

The following two subsections provide a streamlined overview of our technical approach and how we apply it to obtain new dynamic lower bounds, 
as well as discussion and comparison to previous related work.



\subsection{Techniques} \label{subsec_techniques}

To better understand the challenge involved in proving super-logarithmic lower bounds for boolean data structure problems, and 
how our approach departs from previous techniques that fail to overcome it,   
we first revisit Larsen's $\tilde{\Omega}(\lg^2 n)$ lower bound technique for problems with \emph{$\Theta(\log n)$}-bit output size, 
which is most relevant for our work. (We caution that a few 
variations~\cite{CGL15,weinstein:lbs} of Larsen's~\cite{Larsen12a} approach have been proposed, yet all of them crucially rely on large query output size). 
The following overview is presented in the context of 
the \emph{2D range sum} problem for which Larsen originally proved his lower bound. 2D range sum is the 
variant of 2D range counting where each point is assigned a $\Theta(\lg n)$-bit integer weight, and the goal is to return the sum of weights 
assigned to points dominated by the query $q$. Clearly this is a harder problem than 2D range counting (which corresponds to all weights being $1$) and 2D range parity (which again has all weights being $1$, but now only $1$ bit of the output must be returned).

\paragraph{Larsen's Lower Bound~\cite{Larsen12a}.} 
Larsen's result combines the seminal \emph{chronogram method} of Fredman and Saks~\cite{FS89} together with 
the \emph{cell sampling} technique introduced by Panigrahy et al.~\cite{PTW10}. 
The idea is to show that, after $n$ random updates have been 
performed,\footnote{Each update inserts a random point and assigns it a random $\Theta(\lg n)$-bit weight.}
any data structure (with $\poly\log n$ update time) 
must probe many cells when prompted on a random range query.  
To this end, the $n$ random updates are partitioned into $\ell := \Theta(\log n/\log\log n)$ \emph{epochs} $\bU_\ell, \ldots, \bU_i,\ldots, \bU_1$, 
where the $i$-th epoch $\bU_i$ consists of $\beta^i$ updates for $\beta=\mathrm{poly}\log n$. The goal is to show that, for each epoch $i \in \{1,\ldots , \ell \}$, 
a random query must read in expectation $\Omega(\log n/\log\log n)$ 
memory cells whose \emph{last modification} occurred during the $i$th epoch $\bU_i$. 
Summing over all epochs then yields a $\tilde{\Omega}(\log^2 n)$ query lower bound. 

To carry out this approach, one restricts the attention to epoch $i$, assuming all remaining updates in other epochs ($\bU_{-i}$) are 
fixed  (i.e., only $\bU_i$ is random). For a data structure $D$, let $A_i$ denote  the set of memory cells \emph{associated} with epoch $i$, i.e., the cells whose 
\emph{last} update occurred in epoch $i$.  
Clearly, any cell that is written \emph{before} epoch $i$ cannot contain any information about $\bU_i$,
while the construction guarantees there are relatively few cells written \emph{after} epoch $i$, due to the geometric decay in the lengths of epochs.  
Thus, ``most" of the information $D$ 
provides on  
$\bU_i$ comes from cell probes to $A_i$ (hence, intuitively, the chronogram method reduces a dynamic problem into $\approx \log n$ 
nearly independent \emph{static} problems). 

The high-level idea is to now prove that, if a too-good-to-be-true data structure $D$ exists, 
which probes $o(\log n/\log\log n)$ cells associated with epoch $i$ on an average query, 
then $D$ can be used to devise a \emph{compression} scheme (i.e., a ``one-way" communication protocol) which allows 
a decoder to reconstruct the random update sequence $\bU_i$ from an $o(H(\bU_i))$-bit message, an information-theoretic contradiction. 

Larsen's encoding scheme has the encoder (Alice) 
find a subset $C\subseteq A_i$ of a \emph{fixed} size, such that \emph{sufficiently many} range queries $q\in [n] \times [n]$ can be \emph{resolved by $C$}, 
meaning that these queries can be answered without probing any cell in $A_i\setminus C$.
Indeed, the assumption that the query algorithm of $D$ probes only $o(\log n/\log\log n)$ cells from $A_i$,  implies that 
a \emph{random} subset of size 
$|C|=|A_i|/\poly\log n$ 
cells 
resolves at least a $(1/\poly\log n)^{o(\log n/\log\log n)}=n^{-o(1)}$-fraction of the $n^2$ possible queries,
an observation first made in~\cite{PTW10}. This observation in turn implies that by sending the contents and addresses of $C$, the decoder (Bob) can recover the answers to 
some \emph{specific subset} $Q^*\subseteq[n] \times [n]$ of at least $n^{2-o(1)}$ queries.
Intuitively, if the queries of the problem are ``sufficiently independent'', e.g., the answers to all queries are $n$-wise independent over a random $\ep_i$, 
then answering $Q^*$ or even any subset of $Q^*$ of size $n$ would be sufficient to reconstruct the entire update sequence $\ep_i$.
Thus, by simulating the query algorithm $\forall q\in Q^*$ and using the set $C$ to ``fill in" his missing memory cells associated with $\bU_i$, Bob could essentially recover $\bU_i$.
On the other hand, the update sequence itself contains at least $\Omega(|\bU_i|)\gg |C|\cdot w$ bits of entropy, 
hence it cannot possibly be reconstructed from $C$, yielding an information-theoretic contradiction. Here, and throughout the paper, $w$ denotes the number of bits in a memory cell. We make the standard assumption that $w = \Omega(\lg n)$, such that a cell has enough bits to store an index into the sequence of updates performed.

It is noteworthy that range queries do not directly possess such ``$n$-wise independence" property per-se, but using (nontrivial) technical 
manipulations (a-la~\cite{Patrascu07,Larsen12a,weinstein:lbs}) this argument can be made to work, see the discussion in Section~\ref{sec:lowerparity}.

Alas, a subtle but crucial issue with the above scheme is that \emph{Bob cannot identify the subset $Q^*$}, that is, when simulating the query algorithm of 
$D$ on a given query, he can never know whether an \emph{unsampled} ($\notin C$) encountered cell in the query-path in fact belongs to $A_i$ or not.  
This issue is also faced by P\v{a}tra\c{s}cu's approach in~\cite{Patrascu07}.
Larsen resolves this excruciating problem by having Alice further send Bob the indices of (a subset of) $Q^*$ that already reveals enough 
information about $\bU_i$ to get a contradiction.
In order to achieve the anticipated contradiction, the problem must therefore guarantee that the answer to a query reveals more information than it takes to specify 
the query itself ($\Theta(\log n)$ bits for 2D range sum).
This is precisely the reason why Larsen's lower bound requires $\Omega(\log n)$-bit weights assigned to each input point, 
whereas for the boolean 2D range parity problem, all bets are off. 


\subsubsection{Our Techniques} 
We develop a new lower bound technique which ultimately circumvents the aforementioned obstacle that stems from Bob's inability to identify the subset $Q^*$. 
Our high-level strategy is 
 to argue that an efficient dynamic data structure for a boolean problem, induces an efficient one-way protocol 
from Alice (holding the entire update sequence $\cU := \bU_\ell,\ldots,\bU_1$ as before) 
to Bob (who now receives a query $q \in \cQ$ and $\cU\setminus\{\bU_i\}$), which 
enables Bob to answer his 
boolean query with some tiny yet nontrivial advantage over random guessing. 
For a dynamic boolean data structure problem $\cP$, we denote this induced communication game (corresponding to the $i$th epoch) by $G^i_\cP$. 
The following ``weak simulation" theorem, which is the centerpiece of this paper, applies to \emph{any} dynamic boolean data structure 
problem $\cP$:

\begin{theorem}[One-Way Weak Simulation Theorem, informal]	\label{thm_weak_simulation_informal}
Let $\cP$ be any dynamic boolean data structure problem, with $n$ random updates grouped into epochs $\; \cU = \{\ep_i\}_{i=1}^\ell$ 
followed by a single query $q\in \cQ$. 
If $\cP$ admits a dynamic data structure $D$ with word-size $w$, worst-case update time $t_u$ and average (over $\cQ$) expected query time 
$t_q$ with respect to $\cU$, satisfying $t_q,t_u,w\leq n^{0.1}$, then there exists some epoch $i \in [\ell]$ for which there is 
a \emph{one-way} randomized communication protocol for $G^i_\cP$
in which Alice sends Bob a message of only $|\bU_i|/(w t_u)^{\Theta(1)}$ bits, and after which Bob successfully computes $\cP(q,\cU)$ 
with probability at least $1/2 + \exp\left(-t_q \log^2 (w\cdot t_u)/\sqrt{\log n}\right)$.\footnote{Throughout the paper, we use $\exp(x)$ to denote $2^{\Theta(x)}$.}
\end{theorem}

The formal statement and proof of the above theorem can be found in Section~\ref{sec:simulation}.
Before we elaborate on the proof of Theorem~\ref{thm_weak_simulation_informal}, let us explain informally why such a seemingly 
modest guarantee suffices to prove super-logarithmic cell probe lower bounds on boolean problems with a certain ``list-decoding" property.
If we view query-answering as mapping an update sequence to an answer vector,\footnote{An answer vector is a $|\cQ|$-dimensional vector 
containing one coordinate per query, whose value is the answer to this query.} then answering a random query correctly with probability $1/2+e^{-r(n)}$ would correspond to mapping an update sequence to an answer vector that is $(1/2-e^{-r(n)})$-far from the true answer vector defined by the problem.
Intuitively, if the correct mapping defined by the problem is \emph{list-decodable} in the sense that in the $(1/2-e^{-r(n)})$-ball centered at \emph{any} answer vector, there are very few \emph{codewords} (which are the correct answer vectors corresponding to some update sequences), then knowing any vector within distance $(1/2-e^{-r(n)})$ from the correct 
answer vector would reveal a lot of information about the update sequence.
Standard probabilistic arguments~\cite{TCS-010} show that when the \emph{code rate} is $n^{-\Theta(1)}$ (i.e., $|\cQ|=n^{\Theta(1)}$ as for 2D range parity), a random code is ``sufficiently'' list-decodable with $r(n)=\Omega(\log n)$, i.e., for most data structure problems, the protocol in the theorem would reveal too much information if Bob can predict the answer with probability, say $1/2+e^{-0.01\log n}$.
Therefore, Theorem~\ref{thm_weak_simulation_informal} would imply that the query time must be at least $t_q = \Omega(\frac{\log^{1.5} n}{\log^2(w\cdot t_u)})$.
Assuming the data structure has $t_u=\poly\lg n$ worst-case update time and standard word-size $w=\Theta(\lg n)$, 
the above bound gives $t_q \geq \tilde\Omega(\log^{1.5} n)$. 
Indeed all our concrete lower bounds are obtained by showing a similar list-decoding property with $r(n)=\Omega(\log n)$, yielding a 
lower bound of $\tilde\Omega(\log^{1.5} n)$. See Subsection~\ref{sec:concreteproblems} for more details.

\paragraph{Overview of Theorem~\ref{thm_weak_simulation_informal} and the ``Peak-to-Average" Lemma.}
We now present a streamlined overview of the technical approach and proof of our weak one-way simulation theorem, the main result of this paper. 
Let $\cP$ be any boolean dynamic data structure problem and denote by $n_i := |\ep_i| = \beta^i$ the size of each epoch of random updates
(where $\beta := (t_u\cdot w)^{\Theta(1)}$ and  $\sum_{i=1}^\ell n_i = n$).  
Recall that in $G^i_\cP$, Alice receives the entire sequence of epochs $\cU$, Bob receives $q\in_R \cQ$ and $\cU\setminus\{\bU_i\}$, and 
our objective is to show that Alice can send Bob a relatively short message ($n_i/(t_u\cdot w)^{\Theta(1)}$ bits) 
which allows him to compute the answer to $q$ w.r.t $\cU$, denoted $\cP(q,\cU)\in\{0,1\}$, with advantage $\delta := \exp(-t_q\log^2 (w\cdot t_u)/\sqrt{\log n})$ over $1/2$. 

Suppose $\cP$ admits a dynamic data structure $D$ with worst-case update time $t_u$ and expected query time $t_q$ with respect to 
$\cU$ and $q\in_R \cQ$. Following Larsen's cell sampling approach, a natural course of action for Alice is to generate the updated memory 
state $M$ of $D$ (w.r.t $\cU$), and send Bob a relatively small random subset $C_0$ of the \emph{the cells $A_i$ associated with epoch $i$},  
where each cell is sampled with probability $p = 1/(t_u\cdot w)^{\Theta(1)}$.  
Since the expected query time of $D$ is $t_q$ and there are $\ell=\Theta(\log_\beta n)$ epochs, the average (over $i\in [\ell]$) 
number of cells in $A_i$ probed by a query 
is $t_q/\ell$, hence 
the probability that Alice's random set $C_0$ \emph{resolves} Bob's random query $q \in_R \cQ$ is at least $\eps :=p^{\Theta(t_q/\ell)}$. 
Let us henceforth denote this desirable event by $\cW_q$. 
It is easy to see that, if Alice further sends Bob all cells that were written (associated) with \emph{future} epochs $\ep_{<i}$ (which can be done 
using less than $n_i/(w\cdot t_u)^{\Theta(1)}$ bits due to the geometric decay of epochs and the assumption that $D$ probes at most 
$t_u$ cells on each update operation), then \emph{conditioned on $\cW_q$}, Bob would have acquired all the necessary information to perfectly simulate 
the correct query-path of $D$ on his query $q$. 

Thus, if Bob could \emph{detect} the event $\cW_q$, the above argument would have already yielded 
an advantage of roughly $\Pr[\cW_q] \geq \eps = p^{\Theta(t_q/\ell)}\geq \exp(-t_q\log^2 (w\cdot t_u)/\log n) \gg \delta $ (as Bob could simply 
output a random coin-toss unless $\cW_q$ occurs), and this would have finished the proof. 
Unfortunately, certifying the occurrence of $\cW_q$ is prohibitively expensive, precisely for the same reason that identifying the subset $Q^*$ is costly  
in Larsen's argument.
Abandoning the hope for \emph{certifying} the event $\cW_q$ (while insisting on low communication) means that we must take a 
fundamentally different approach to argue that the noticeable occurrence of this event can somehow 
still be exploited implicitly so as to guarantee a nontrivial 
advantage. This is the heart of the paper, 
and the focal point of the rest of this exposition. 

The most general strategy Bob has is to output his ``maximum likelihood" estimate for the answer $\cP(q,\cU)$ given the information he receives, 
i.e., the more likely posterior value of $(\cP(q,\cU) \; | \ep_{-i},C_0) \in \{0,1\}$ (for simplicity of exposition, we henceforth ignore the conditioning 
on $\ep_{-i},C_0$ and on the set of updates $D$ makes to future epochs $\ep_{<i}$ which Alice sends as well). 
Assuming without loss of generality that the answer to the query is $\cP(q,\cU)=1$, when $\cW_q$ occurs, this strategy produces an advantage (``bias") 
of $1/2$ (since when $\cW_q$ occurs, the answer $\cP(q,\cU)$ is completely determined by $\ep_{-i},C_0$ and the updates to $\ep_{<i}$), 
and when it does not occur, the strategy produces a bias of $\Pr[(\cP(q,\cU)=1 |\overline{\cW_q})] - 1/2$.
Thus, the overall bias is $\Pr[\cW_q]\cdot (1/2) + \Pr[\overline{\cW_q}]\cdot\left( \Pr[(\cP(q,\cU)=1 |\overline{\cW_q})] - 1/2\right)$.
This quantity could be arbitrarily close to $0$, since we have no control over the 
distribution of the answer conditioned on the complement event $\overline{\cW_q}$, which might even cause perfect cancellation of the two terms.

Nevertheless, one could hope that such unfortunate cancellation of our advantage can be avoided if Alice reveals to Bob some little extra ``relevant" information.  
To be more precise, let $S_q$ be the set of memory addresses $D$ would have probed when invoked on the query $q$ \emph{according to Bob's simulation}.
That is, Bob simulates $D$ until epoch $i$, updates the contents for all cells that appear in Alice's message, and simulates the query algorithm for $q$ on this memory state.
In particular, \emph{if} the event $\cW_q$ occurs, then $S_q$ is the correct  set of memory cells the data structure probes.
Of course, the set $S_q$ is extremely \emph{unlikely} to be ``correct'' as $\Pr[\cW_q]$ is tiny, so $S_q$ should generally be viewed as an arbitrary subset of memory addresses.
Now, the \emph{true contents} of the cells $S_q$ (w.r.t the true memory state $M$) induce some \emph{posterior distribution} on the \emph{correct} 
answer $\cP(q,\cU)$ 
(in particular, when $\cW_q$ occurs, the path is correct and its contents induce the true answer).

Imagine that Alice further reveals to  Bob the true contents of some small subset $Y\subseteq S_q$, i.e., an assignment $x\in [2^w]^Y$.
The posterior distribution of the answer $\cP(q,\cU)$ conditioned on $x$ is simply the convex combination of the posterior 
distributions conditioned on ``$S_q=z$'' for all $z$'s that are consistent with $x$ ($z|_Y=x$), weighted by the probability of $z$ ($\Pr[S_q=z]$) up to some normalizer.
The contribution of each term in this convex combination (i.e., of each posterior distribution induced by a partial assignment $x$) to the overall bias,  
is precisely the average, over all \emph{full} assignments $z$ to cells in $S_q$ which are $x$-consistent, 
of the posterior bias induced by the event ``$S_q=z$'' (i.e., when the entire $S_q$ is revealed).
For each full assignment $z$, we denote its latter contribution by $f(z)$, hence 
 the expected bias contributed by the event ``$z|_Y=x$'' is nothing but the sum of $f(z)$ over all $z$'s satisfying $z|_Y=x$.
Furthermore, we know that there is some assignment $z^*$, namely the contents of $S_q$ when $\cW_q$ occurs, such that $|f(z^*)|$ is ``large'' (recall the bias is $1/2$ in this event). 
Thus, the key question we pose and set out to answer, is whether it is possible to translate this $\ell_\infty$ ``peak''  of $f$ into 
a comparable lower bound on the \emph{``average''} bias $\sum_{x}\left|\sum_{z|_Y=x} f(z)\right|$, by conditioning on the assignments to a small subset of coordinates $Y$.
Indeed, if such $Y$ exists, Alice can sample independently another set of memory cells $C_1$ and send it to Bob.
With probability $p^{|Y|}$, all contents of $Y$ are revealed to Bob, and we will have the desired advantage.
In essence, the above question is equivalent to the following information-theoretic problem:

\begin{displayquote}
\emph{
Let $Z$ be a $k$-variate random variable and $B$ a uniform binary random variable in the same probability space, satisfying:
(i) $\Pr[Z=z^*]\geq \eps$ for some $z^*$; \; (ii)  $H(B\mid Z=z^*)=0$. 
What is the smallest subset of coordinates $Y\subseteq [k]$ such that 
	$H(B\mid (Z|_Y))\leq 1-\eta$ ?
}
\end{displayquote}

The crux of our proof is the following lemma, which asserts that conditioning on only $|Y| = O(\sqrt{k\log(1/\eps)})$ many coordinates suffices 
 to achieve a non-negligible \emph{average} advantage $\eta= \exp(-\sqrt{k\log(1/\eps)})$.  

\begin{lemma}[Peak-to-Average Lemma]\label{lem_peak_to_avg}
Let $f : \Sigma^k \to \R $ be any real function 
satisfying: (i) $\sum_{z\in\Sigma^k}|f(z)|\leq 1$; and (ii) $\max_{z\in\Sigma^k}|f(z)| \geq \eps $. 
Then there exists a subset $Y$ of indices, $|Y| \leq O\left(\sqrt{k\cdot \log 1/\eps}\right)$, such that 
$\sum_{y\in\Sigma^Y}|\sum_{z|_Y=y}f(z)|\geq\exp(-\sqrt{k\cdot \log 1/\eps})$. 
\end{lemma} 

An indispensable ingredient of the proof is the usage of low-degree (multivariate) polynomials with ``threshold"-like phenomena, commonly known as 
\emph{(discrete) Chebyshev polynomials}.\footnote{These are real polynomials defined on the $k$-hypercube, of degree $O(\sqrt{k\log(1/\gamma)})$ 
and whose value is uniformly bounded by $\gamma$ everywhere \emph{on the cube} except the all-$0$ point which attains the value $1$.} 
The lemma can be viewed as an interesting and efficient way of ``decomposing" a distribution into a small number of conditional distributions, 
``boosting" the effect of a single desirable event, hence the Peak-to-Average Lemma may be of independent interest  
(see Section~\ref{subsec_PTA_lemma} for a high-level overview and the formal proof). 
In Appendix \ref{app_pf_tight_PtA}, we show that the lemma is in fact tight, in the sense that there are functions for which conditioning on $o\left(\sqrt{k\cdot \log 1/\eps}\right)$ 
of their coordinates provides \emph{no advantage at all}.

To complete the proof of the simulation theorem, we apply the Peak-to-Average Lemma with $f$, $k:=t_q$ and $\eps := p^{\Theta(t_q/\ell)} = (1/wt_u)^{O(t_q/\ell)}$. 
The lemma guarantees that Bob can find a  small (specific) set of coordinates $Y\subseteq S_q$, such that  
his maximum-likelihood estimate 
conditioned on the true value $y$ of the coordinates in $Y$ 
must provide an advantage of at least $\exp(-\sqrt{k\cdot \log 1/\eps}) =  \exp\left(-t_q \log (w\cdot t_u)/\sqrt{\log n}\right)$. 
Since $|Y|$ is small, the probability that $Y$ is contained in Alice's second sample $C_1$ is $p^{|Y|}\geq \exp\left(-t_q \log^2 (w\cdot t_u)/\sqrt{\log n}\right)$.
Overall, Bob's maximum-likelihood strategy 
provides the desired advantage $\delta$ we sought. 



\subsection{Applications: New Lower Bounds}
\label{sec:concreteproblems}
We apply our new technique to a number of classic data structure
problems, resulting in a range of new lower bounds. This section
describes the problems and the lower bounds we derive for them, in context of prior work. As a
warm-up, we prove a lower bound for a somewhat artificial version of
polynomial evaluation:

\paragraph{Polynomial Evaluation.}
Consider storing, updating and evaluating a polynomial $P$ over the Galois
field $\GF(2^d)$. Here we assume that elements of $\GF(2^d)$ are
represented by bit strings in $\{0,1\}^d$, i.e. there is some
bijection between $\GF(2^d)$ and $\{0,1\}^d$. Elements are represented
by the corresponding bit strings. Any bijection between elements and
bit strings suffice for our lower bound to apply.

The \emph{least-bit polynomial evaluation} data structure problem is defined as
follows: A degree $n \leq 2^{d/4}$ polynomial $P(x) = \sum_{i=0}^n a_i
x^i$ over $\GF(2^d)$ is initialized with all $n+1$ coefficients $a_i$
being $0$. An update is specified by a tuple $(i, b)$ where $i \in
[n+1]$ is an index and $b$ is an element in $\GF(2^d)$. It changes the
coefficient $a_i$ such that $a_i \gets a_i + b$ (where addition is
over $\GF(2^d)$). A query is specified by an element $y \in \GF(2^d)$
and one must return the least significant bit of $P(y)$. Recall that
we make no assumptions on the concrete representation of the elements
in $\GF(2^d)$, only that the elements are in a bijection with
$\{0,1\}^d$ so that precisely half of all elements in $\GF(2^d)$ have
a $0$ as the least significant bit. 

Using our weak one-way simulation theorem, Section~\ref{sec:polyeval}
proves the following lower bound:

\begin{theorem}
\label{thm:polyeval}
Any cell probe data structure for \emph{least-bit polynomial
  evaluation} over $\GF(2^d)$, having cell size $w$, worst case update
time $t_u$ and expected average query time $t_q$ must satisfy:
$$
t_q = \Omega\left( \min\left\{ \frac{d \sqrt{\lg n} }{\lg^2(t_u w)}, \frac{\sqrt{n}}{(t_u w)^{O(1)}} \right\}\right).
$$
\end{theorem}

Note that this lower bound is not restricted to have $d = O(\lg n)$ (corresponding to having polynomially many queries). It holds for arbitrarily large $d$ and thus demonstrates that our lower bound actually grows as log of the number of queries, times a $\sqrt{\lg n}$. At least up to a certain (unavoidable) barrier (the $\sqrt{n}$ bound in the min is precisely when the query time is large enough that the data structure can read all cells associated to more than half of the epochs). We remark that the majority of previous lower bound techniques could also replace a $\log n$ in the lower bounds by a $d$ for problems with $2^d$ queries. Our introduction focuses on the most natural case of polynomially many queries ($d = \Theta(\lg n)$) for ease of exposition.
   
Polynomial evaluation has been studied quite intensively from a lower
bound perspective, partly since it often allows for very clean proofs. The previous work on the problem considered the standard (non-boolean)  
version in which we are required to output the value $P(x)$, not just its 
least significant bit. Miltersen~\cite{Miltersen:polyn} first considered the static
version where the polynomial is given in advance and we disallow
updates. He proved a lower bound of $t_q = \Omega(d/\lg S)$ where $S$
is the space usage of the data structure in number of cells. This was
improved by Larsen~\cite{Larsen:2012:focs} to $t_q = \Omega(d/\lg(Sw/(nd)))$, which remains
the highest static lower bound proved to date. Note that the lower bound peaks at $t_q = \Omega(d)$ for linear space $S=O(nd/w)$. Larsen~\cite{Larsen:2012:focs} also
extended his lower bound to the dynamic case (though for a slightly
different type of updates), resulting in a lower bound of $t_q =
\Omega(d \lg n/(\lg(w t_u) \cdot \lg(w t_u/d))$. Note that none of these lower
bounds are greater than $t_q = \Omega(\lg n/\lg t_u)$ per output bit
and in that sense they are much weaker than our new lower bound.

In~\cite{gal:succinct}, G\'{a}l and Miltersen
considered \emph{succinct} data structures for polynomial
evaluation. Succinct data structures are data structures that use
space close to the information theoretic minimum required for storing
the input. In this setting, they showed that any data structure for
polynomial evaluation must satisfy $t_q r = \Omega(n)$ when $2^d \geq
(1+\eps)n$ for any constant $\eps >0$. Here $r$ is the
\emph{redundancy}, i.e. the \emph{additive} number of extra bits of
space used by the data structure compared to the information theoretic
minimum. Note that even for data structures using just a factor $2$ more space than the minimum possible, the time lower bound reduces to the trivial $t_q = \Omega(1)$. For data structures with \emph{non-determinism}
(i.e., they can guess the right cells to probe), Yin~\cite{yin:nondeterm} proved a lower bound matching that of
Miltersen.

On the upper bound side, Kedlaya and Umans~\cite{kedlaya:polyeval} presented a
word-RAM data structure for the static version of the problem, having
space usage $n^{1+\eps} d^{1+o(1)}$ and worst case query time $\lg^{O(1)} n \cdot
d^{1+o(1)}$, getting rather close to the lower bounds. While not
discussed in their paper, a simple application of the
\emph{logarithmic method} makes their data structure dynamic with an
amortized update time of $n^{\eps} d^{1+o(1)}$ and worst case query time $\lg^{O(1)} n \cdot
d^{1+o(1)}$.


\paragraph{Parity Searching in Butterfly Graphs.}
In a seminal paper~\cite{patrascu08structures}, P{\v a}tra{\c s}cu
presented an exciting connection between an entire class of data
structure problems. Starting from a problem of \emph{reachability
  oracles in the Butterfly graph}, he gave a series of reductions to
classic data structure problems. His reductions resulted in $t_q = \Omega(\lg
n/\lg(Sw/n))$ lower bounds for static data structures solving any of
these problems. 

We modify P{\v a}tra{\c s}cu's reachability problem
such that we can use it in reductions to prove new \emph{dynamic} lower
bounds. In our version of the problem, which we term \emph{parity searching in Butterfly graphs},  
the data structure must maintain a set of directed acyclic graphs (Butterfly graphs of the same degree $B$, but different depths) under updates which assign binary weights to edges, 
and support queries that ask to compute the parity of weights assigned to edges along a number of paths in these graphs. The formal definition of this version of the problem 
is deferred to Section~\ref{sec:paritysearchdef}.  

While this new problem might sound quite artificial and incompatible to work with, we 
show that parity searching in Butterfly graphs in fact 
reduces to many classic problems, hence 
proving lower bounds on this problem is the key to many of our results. 
Indeed, our starting point is the following lower bound: 

\begin{theorem}
\label{thm:paritysearch}
Any dynamic data structure for \emph{parity searching in Butterfly graphs} of degree $B = (w t_u)^8$, with a total of $n$ edges, having cell size $w$, worst case update
time $t_u$ and expected average query time $t_q$ must satisfy:
$$
t_q = \Omega\left(\frac{\lg^{3/2} n}{\lg^3(t_u w)}\right).
$$
\end{theorem}

In the remainder of this section, we present new lower bounds which we 
derive via reductions from parity searching in Butterfly graphs . For context, our results are complemented 
with a discussion of  previous work. 

\paragraph{2D Range Counting.}
In 2D range counting, we are given $n$ points $P$ on a
$[U]\times[U]$ integer grid, for some $U = n^{O(1)}$. We must
preprocess the points such that given a query point $q=(x,y) \in [U]
\times [U]$, we can return the number of points $p \in P$ that are
\emph{dominated} by $q$ (i.e. $p.x \leq q.x$ and $p.y \leq q.y$). In the
dynamic version of the problem, an update specifies a new point to
insert. 2D range counting is a fundamental problem in both computational
geometry and spatial databases and many variations of it have been
studied over the past many decades.

Via a reduction from reachability oracles in the Butterfly graph, P{\v
  a}tra{\c s}cu~\cite{patrascu08structures} proved a static lower bound of $t_q = \Omega(\lg
n/\lg(Sw/n))$ for this problem, even in the case where one needs only to 
return the \emph{parity} of the number of points dominated by $q$. Recall that this is the \emph{2D range parity} problem.

It turns out that a fairly easy adaptation of P{\v a}tra{\c s}cu's reduction 
implies the following:

\begin{theorem}
\label{thm:reducerangeparity}
Any dynamic cell probe data structure for 2D range parity, having cell size $w$, worst case update
time $t_u$ and expected query time $t_q$, gives a
dynamic cell probe data structure for parity searching in Butterfly
graphs (for any degree $B$) with cell size $w$, worst case update time $O(t_u)$ and average expected query time $t_q$.
\end{theorem}

Combining this with our lower bound for parity searching in
Butterfly graphs (Theorem~\ref{thm:paritysearch}), we obtain:

\begin{corollary}\label{thm_orc}
Any cell probe data structure for 2D range parity, having cell size $w$, worst case update
time $t_u$ and expected query time $t_q$ must satisfy:
$$
t_q = \Omega\left(\frac{\lg^{3/2} n}{\lg^3(t_u w)}\right).
$$
\end{corollary}

In addition to P{\v a}tra{\c s}cu's static lower bound, Larsen~\cite{Larsen12a}
studied the aforementioned variant of the range counting problem, called \emph{2D range sum}, in which 
points are assigned $\Theta(\lg n)$-bit integer weights and the goal is to compute the sum of weights assigned
to points dominated by $q$. 
As previously discussed, Larsen's lower bound for dynamic 2D range sum was $t_q =
\Omega((\lg n/ \lg(t_u w))^2)$ and was the first lower bound to break
the $\Omega(\lg n)$-barrier, though only for a problem with $\Theta(\lg
n)$ bit output. Weinstein and Yu~\cite{weinstein:lbs} later re-proved
Larsen's lower bound, this time extending it to the setting of
amortized update time and a very high probability of error. Note that
these lower bounds remain below the logarithmic barrier when measured
per output bit of a query. While 2D range counting (not the parity
version) also has $\Theta(\lg n)$-bit outputs, it seems that the
techniques of Larsen and Weinstein and Yu are incapable of proving an
$\omega(\lg n)$ lower bound for it. Thus the strongest previous lower bound for
the dynamic version of 2D range counting is just the static bound of
$t_q = \Omega(\lg n/ \lg(t_u w))$ (since one cannot build a data
structure with space usage higher than $S=t_u n$ in $n$ operations). As a rather technical explanation
for why the previous techniques fail, it can be observed that they all argue that
a collection of $m=n/\poly(\lg n)$ queries have $\Omega(m \lg n)$ bits
of entropy in their output. But for 2D range counting, having
$n/\poly(\lg n)$ queries means that on average, each query contains
just $\poly(\lg n)$ new points, reducing the total entropy to
something closer to $O(m \lg \lg n)$. This turns out to be useless for
the lower bound arguments. It is conceivable that a clever argument could show that the entropy remains $\Omega(m \lg n)$, but this has so forth resisted all attempts.

From the upper bound side, J{\'a}J{\'a}, Mortensen and Shi~\cite{jaja:counting} gave a
static 2D range counting data structure using linear space and
$O(\lg n/\lg \lg n)$ query time, which is optimal by
P{\v a}tra{\c s}cu's lower bound. For the dynamic case, Brodal et
al.~\cite{brodal:rangemedian} gave a data structure with $t_q = t_u = O((\lg n/\lg \lg
n)^2)$. Our new lower bound shrinks the gap between the upper and
lower bound on $t_q$ to only a factor $\sqrt{\lg n} \lg \lg n$ for $t_u = \poly(\lg n)$.

\paragraph{2D Rectangle Stabbing.}
In 2D rectangle stabbing, we must maintain a set of $n$ 2D axis aligned
rectangles with integer coordinates, i.e. rectangles are of the form
$[x_1 , x_2] \times [y_1, y_2]$. We assume coordinates are bounded by
a polynomial in $n$. An update inserts a new rectangle. A query is specified by a point $q$, and one must
return the number of rectangles containing $q$. This problem is known
to be equivalent to 2D range counting via a folklore reduction. Thus
all the bounds in the previous section, both upper and lower bounds,
also apply to this problem. Furthermore, 2D range parity is also equivalent to 2D rectangle parity, i.e. returning just the parity of the number of rectangles stabbed.

\paragraph{Range Selection and Range Median.}
In range selection, we are to store an array $A =
\{A[0],\dots,A[n-1]\}$ where each entry stores an integer bounded by a
polynomial in $n$. A query is specified by a triple $(i,j,k)$. The goal is to return the index of the $k$'th smallest entry
in the subarray $\{A[i],\dots,A[j]\}$. In the dynamic version of the
problem, entries are initialized to $0$. Updates are specified by an index $i$ and a value $a$ and has
the effect of changing the value stored in entry $A[i]$ to $a$. In
case of multiple entries storing the same value, we allow returning an
arbitrary index being tied for $k$'th smallest.

We give a reduction from parity searching in Butterfly graphs:
\begin{theorem}
\label{thm:reduceselect}
Any dynamic cell probe data structure for range selection, having cell size $w$, worst case update
time $t_u$ and expected query time $t_q$, gives a
dynamic cell probe data structure for parity searching in Butterfly
graphs (for any degree $B$) having cell size $w$, worst case update time $O(t_u \lg^2 n)$ and expected average query time $t_q$. Furthermore, this holds even if we force $i=0$ in queries $(i,j,k)$ and require only that we return whether the $k$'th smallest element in $A[0],\dots,A[j]$ is stored at an even or odd position.
\end{theorem}

Since we assume $w = \Omega(\lg n)$, combining this with Theorem~\ref{thm:paritysearch} immediately proves the following:

\begin{corollary}
\label{thm:rangeselect}
Any cell probe data structure for range selection, having cell size $w$, worst case update
time $t_u$ and expected query time $t_q$ must satisfy:
$$
t_q = \Omega\left(\frac{\lg^{3/2} n}{\lg^3(t_u w)}\right).
$$
Furthermore, this holds even if we force $i=0$ in queries $(i,j,k)$ and require only that we return whether the $k$'th smallest element in $A[0],\dots,A[j]$ is stored at an even or odd position.
\end{corollary}

While range selection is not a boolean data structure problem, it is
still a fundamental problem and for the same reasons as mentioned
under 2D range counting, the previous lower bound techniques seem
incapable of proving $\omega(\lg n)$  lower bounds for the dynamic
version. Thus we find our new lower bound very valuable despite the
problem not beeing boolean . Also, we do in fact manage to prove the same lower bound for the boolean version where we need only determine whether the index of the $k$'th smallest element is even or odd.

For the static version of the problem, J\o rgensen and Larsen~\cite{larsen:median}
proved a lower bound of $t_q = \Omega(\lg n/\lg(Sw/n))$. Their proof
was rather technical and a new contribution of our work is that their
static lower bound now follows by reduction also from  P{\v a}tra{\c
  s}cu's lower bound for reachability oracles in the Butterfly
graph. For the dynamic version of the problem, no lower bound stronger than the $t_q=\Omega(\lg n/\lg(t_u w))$ bound following from the static bound was previously known.

On the upper bound side, Brodal et
al.~\cite{brodal:rangemedian} gave a linear space static data
structure with query time $t_q = O(\lg n/\lg \lg n)$. This matches the
lower bound of J\o rgensen and Larsen. They also gave a dynamic data structure with $t_q = t_u = O((\lg n/\lg \lg
n)^2)$. 

Since we prove our lower bound for the version of range
selection where $i=0$, also known as prefix selection, we can re-execute a
reduction of J\o rgensen and Larsen~\cite{larsen:median}. This means that we also get a
lower bound for the fundamental range median problem. Range median is the natural special
case of range selection where $k = \lceil (j-i+1)/2\rceil$. 

\begin{corollary}
Any cell probe data structure for range median, having cell size $w$, worst case update
time $t_u$ and expected query time $t_q$ must satisfy:
$$
t_q = \Omega\left(\frac{\lg^{3/2} n}{\lg^3(t_u w)}\right).
$$
Furthermore, this holds even if we are required only to return whether the median amongst $A[i],\dots,A[j]$ is stored at an even or odd position.
\end{corollary}
We note that the upper bound of Brodal et al. for range selection is also the best known 
upper bound for range median.

\section{Organization} 
In Section~\ref{sec:prelim} we introduce both the dynamic cell probe model and  
the one-way communication model, which is the main proxy for our results. 
In Section~\ref{sec:simulation} we state the formal version of Theorem~\ref{thm_weak_simulation_informal} and give its proof as well as the proof 
of the Peak-to-Average lemma. Section~\ref{sec:polyeval} and onwards are devoted to applications of our new simulation theorem, starting with 
a lower bound for polynomial evaluation. In Section~\ref{sec:lowerparity} we formally define parity searching in Butterfly graphs and prove a lower 
bound for it using our simulation theorem. 
Finally, Section~\ref{sec:reductions} presents a number of reductions from parity searching in Butterfly graphs to various fundamental data structure problems, 
proving the remaining lower bounds stated in the introduction. 

\ifx\mainfile\undefined
\bibliographystyle{alpha}
\bibliography{refs}
\end{document}
\fi

\section{Preliminaries}
\label{sec:prelim}
\paragraph{The dynamic cell probe model.}
{A dynamic data structure in the cell probe model consists of an array of memory cells, each of which can store $w$ bits. 
Each memory cell is identified by a $w$-bit address, so the set of possible addresses is $[2^w]$. It is natural to assume that 
each cell has enough space to address (index) all update operations performed on it, hence we assume that 
$w=\Omega(\log n)$ when analyzing a sequence of $n$ operations. 

Upon an update operation, the data structure can perform read and write operations to its memory so as to reflect the update, 
by \emph{probing} a subset of memory cells. This subset may be an arbitrary function of the update and the content of the memory 
cells previously probed during this process. The update time of a data structure, denoted by $t_u$, is the number of probes 
made when processing an update (this complexity measure can be measured in worst-case or in an amortized sense). 
Similarly,  upon a query operation, the data structure performs a sequence of probes to read a subset of the memory cells in order 
to answer the query. Once again, this subset may by an arbitrary (adaptive) function of the query and previous cells probed during the 
processing of the query. The query time of a data structure, denoted by $t_q$, is the number of probes made when processing a query.}

\subsection{One-way protocols and ``Epoch" communication games} 
A useful way to abstract the information-theoretic bottleneck of
dynamic data structures is \emph{communication complexity}. 
Our main results (both upper and lower bounds) are cast in terms of the following  two-party communication games, which are induced by 
dynamic data structure problems:

\begin{definition}[Epoch Communication Games $G^i_\cP$]
Let $\cP$ be a dynamic data structure problem, consisting of a sequence of $n$ update operations divided into \emph{epochs}  
$\cU=(\ep_\ell,\ep_{l-1},\ldots,\ep_1)$, where $|\ep_i|=n_i$ (and $\sum_i n_i = n$),  
followed by a single query $q\in \cQ$. For each epoch $i\in[\ell]$, the two-party communication game $G^i_\cP$ induced by $\cP$ 
is defined as follows: 
\begin{itemize}
\item Alice receives all update operations $\cU=(\ep_\ell,\ep_{l-1},\ldots,\ep_1)$. 
\item Bob receives $\ep_{-i} := \cU\setminus \{\ep_i\}$ (i.e., all updates \emph{except} those in epoch $i$) and a query $q\in \cQ$ for $\cP$. 
\item The goal of the players is to output the correct answer to $q$, that is, to output $\cP(q,\cU)$. 
\end{itemize}
\end{definition}

\ 

We shall consider the following restricted model of communication for solving such communication games. 

\begin{definition}[One-Way Randomized Communication Protocols]
Let $f : \cX \times \cY \mapsto \{0,1\}$ be a two-party boolean function. A \emph{one-way} communication protocol $\pi$
for $f(x,y)$ under input distribution $\mu$ proceeds as follows:  
\begin{itemize}
\item Alice and Bob have shared access to a public random string $R$ of their choice. 
\item Alice sends Bob a \emph{single} message, $M_A(x,R)$, which is only a function of her input and the public random string. 
\item Based on Alice's message, Bob must output a value $v_\pi = v_\pi(y, R, M_A) \in \{0,1\}$. 
\end{itemize}

We say that \emph{$\pi$ $\eps$-solves $f$ under $\mu$ with cost $m$}, if :
\begin{itemize}
	\item
		For any input $x$, Alice never sends more than $m$ bits to Bob, i.e., $|M_A(x, R)| \leq m$, for all $x,r$. 
	\item 
		$ \Pr_{(x,y)\sim \mu, R}[v_\pi = f(x,y)] \geq 1/2 + \eps. $ 
\end{itemize}
 
\end{definition}
\vspace{.1in}
Let us denote by 
\[ \suc(f,\mu,m) := \sup\{\eps \mid \text{ $\exists$ one-way protocol $\pi$ that $\eps$-solves $f$ under $\mu$ with cost $m$} \} \] 
the largest advantage $\eps$ achievable for predicting $f$ under $\mu$ via an $m$-bit one-way communication protocol. 
For example, when applied to the boolean communication problem $G^i_\cP$, we say that $G^i_\cP$ has an $m$-bit one-way communication protocol 
with advantage $\eps$, if $\suc(G^i_\cP,\mu,m) \geq \eps$.  We remark that we sometimes use the notation $\|\pi\|$ to denote the message-length (i.e., number of bits $m$) 
of the communication protocol $\pi$.

\ifx\mainfile\undefined
\documentclass{article}

\usepackage{fullpage, amsmath, amsthm, amssymb}
\usepackage{enumerate}
\usepackage{float}
\usepackage{color}
\newtheorem{definition}{Definition}
\newtheorem{theorem}{Theorem}
\newtheorem{corollary}{Corollary}
\newtheorem{lemma}{Lemma}
\newtheorem{proposition}{Proposition}
\newtheorem{claim}{Claim}
\newtheoremstyle{restate}{}{}{\itshape}{}{\bfseries}{~(restated).}{.5em}{\thmnote{#3}}
\theoremstyle{restate}
\newtheorem*{restate}{}

\newcommand{\algwidth}{0.97\textwidth}
\newcommand{\Fam}{\mathcal{F}}
\DeclareMathOperator{\mmod}{mod}
\renewcommand{\P}{\textbf{P}}
\newcommand{\Q}{\textbf{Q}}
\newcommand{\E}{\mathop{{}\mathbb{E}}}
\renewcommand{\S}{\textbf{S}}
\newcommand{\p}{\textbf{p}}
\renewcommand{\log}{\lg}

\newcommand{\R}{\mathbb{R}}
\newcommand{\eps}{\epsilon}
\newcommand{\path}{\mathsf{path}}
\newcommand{\suc}{\overrightarrow{\mathsf{adv}}}

\newcommand{\mnote}[1]{ \marginpar{\tiny\bf
            \begin{minipage}[t]{0.5in}
              \raggedright #1
           \end{minipage}}}
           
\newcommand{\GF}{GF}
\newcommand{\cA}{\mathcal{A}}
\newcommand{\cB}{\mathcal{B}}
\newcommand{\cC}{\mathcal{C}}
\newcommand{\cE}{\mathcal{E}}
\newcommand{\cI}{\mathcal{I}}
\newcommand{\cP}{\mathcal{P}}
\newcommand{\cG}{\mathcal{G}}
\newcommand{\cH}{\mathcal{H}}
\newcommand{\cM}{\mathcal{M}}
\newcommand{\cN}{\mathcal{N}}
\newcommand{\cX}{\mathcal{X}}
\newcommand{\cY}{\mathcal{Y}}
\newcommand{\cS}{\mathcal{S}}
\newcommand{\cQ}{\mathcal{Q}}
\newcommand{\cU}{\mathcal{U}}
\newcommand{\cT}{\mathcal{T}}
\newcommand{\cR}{\mathcal{R}}
\newcommand{\cL}{\mathcal{L}}
\newcommand{\cF}{\mathcal{F}}
\newcommand{\cZ}{\mathcal{Z}}
\newcommand{\cD}{\mathcal{D}}
\newcommand{\cJ}{\mathcal{J}}
\newcommand{\cK}{\mathcal{K}}
\newcommand{\cO}{\mathcal{O}}
\newcommand{\cW}{\mathcal{W}}

\newcommand{\bU}{\mathbf{U}}

\newcommand{\tS}{\tilde{S}}
\newcommand{\ORC}{\oplus\mathsf{ORC}}
\newcommand{\orc}{\mathsf{ORC}}
\newcommand{\poly}{\mathrm{poly}}
\newcommand{\ep}{\mathbf{U}}

\DeclareMathOperator{\supp}{supp}
\DeclareMathOperator{\rank}{rank}

\newcommand{\nz}{non-zero}
\newcommand{\AND}{\mathsf{AND}_k}

\begin{document}

\fi
\section{One-Way Weak Simulation of Dynamic Data Structures}
\label{sec:simulation}

In this section we prove our main result, Theorem \ref{thm_weak_simulation_informal}.  
For \emph{any} dynamic decision problem $\cP$, 
we show that if $\cP$ admits an efficient data structure $D$  
with respect to a random sequence of $n$ updates divided into $\ell:=\log_\beta n$ epochs
$\cU=(\ep_\ell,\ep_{\ell-1},\ldots,\ep_1)$, 
then we can use it to devise an efficient \emph{one-way} communication protocol for the underlying 
two-party communication problem $G^i_\cP$ of \emph{some} (large enough) epoch $i$, with a nontrivial 
success (advantage over random guessing). 

Throughout this section, let us denote the size of epoch $i$ by $n_i := |\ep_i|=\beta^i$, where we require $\beta = (w\cdot t_u)^{\Theta(1)}$, and $\sum_{i=1}^\ell n_i = n$. 
We prove the following theorem.  

\begin{restate}[Theorem \ref{thm_weak_simulation_informal}]
Let $\cP$ be a dynamic boolean data structure problem, with $n$ random updates grouped into epochs $\cU = \{\ep_i\}_{i=1}^\ell$, such that $|\ep_i|=\beta^i$,
followed by a single query $q\in \cQ$. If $\cP$ admits a dynamic data structure $D$ with worst-case update time $t_u$ and 
average (over $\cQ$) expected query time $t_q$ satisfying $t_q(w\cdot t_u)^{a+1}\leq n^{1/2}$, then there exists some epoch $i \in [\ell/2, \ell]$ for which  
\[ \suc\left(G^i_\cP, \;\cU, \; n_i/(w\cdot t_u)^{a-1}\right) \geq  \exp\left(-t_q \log^2 (w\cdot t_u)/\sqrt{\log n}\right) \]
as long as $\beta  =  (w\cdot t_u)^{\Theta(1)} \geq (w\cdot t_u)^a$ for a constant $a>1$.
\end{restate}

\begin{proof}

Consider the memory state $M=M(\cU)$ of $D$ after the entire update sequence $\cU$, and for each cell $c\in M$, define its \emph{associated epoch} $E(c)$ to be  
 the \emph{last} epoch in $[\ell]$ during which $c$ was probed (note that $E(c)$ is a random variable over the random update sequence $\cU$). 
For each query $q\in \cQ$, let $T_q$ be the random variable denoting the number of probes made by $D$ on query $q$ (on the random update sequence). 
For each query $q$ and epoch $i$, let $T_q^i$ denote the number of probes on query $q$ to cells \emph{associated with epoch $i$} (i.e., cells $c$ for 
which $E(c)=i$). 

By definition, we have $\frac{1}{|\cQ|}\sum_{q\in \cQ}\E[T_q]=t_q$ and $T_q=\sum_{i=1}^{\ell} T_q^i$. 
By averaging, there is an epoch $i\in [\ell/2,\ell]$ such that $\frac{1}{|\cQ|}\sum_{q\in \cQ}\E[T_q^i]\leq 2t_q/\ell$. 
By Markov's inequality and a union bound, there exists a subset $\cQ'\subseteq\cQ$ of $|\cQ|/2$ queries such that both
\begin{equation}\label{eq_qury_time_epoch_i}
\E[T^i_q] \leq 8t_q/\ell \;\;\;\; \text{and} \;\;\;\;\;  \E[T_q] \leq 8t_q  \; , 
\end{equation} 
for every query $q\in \cQ'$. 
By Markov's inequality and union bound, for each $q\in \cQ'$, we have
\begin{equation}\label{eq_qtime_ep_i_two}
\Pr_{\cU}[T_q^i\leq 32t_q/\ell,T_q\leq 32 t_q]\geq 1/2.
\end{equation}
Note that, while Bob cannot identify the event ``$T_q^i\leq 32t_q/\ell,T_q\leq 32 t_q$" (as it depends on Alice's input as well), 
he does know whether his query $q$ is in $\cQ'$ or not, which is enough to certify \eqref{eq_qtime_ep_i_two}. 

Now, suppose that Alice samples each cell \emph{associated with epoch $i$} in $M$ independently with probability $p$, where 
\[ p := \frac{1}{\left(w\cdot t_u\right)^{a}}\]
(note that, by definition of $G^i_\cP$, Alice can indeed generate the memory state $M$ and compute the associated epoch for each cell, as her input 
consists of the entire update sequence). Let $C_0$ be the resulting set of cells sampled by Alice. 
Alice sends Bob $C_0$ (both addresses and contents). 
For a query $q\in \cQ'$, let $\cW_q$ denote the event that the set of cells $C_0$ Bob receives, contains \emph{all} $T_q^i$ cells associated with epoch 
$i$ and probed by the data structure. 
By Equation~\eqref{eq_qtime_ep_i_two},
we have that for every $q\in\cQ'$
\begin{equation}\label{eq_pr_all_black_sampled} 
\Pr_{C_0,\cU}[\cW_q,T_q\leq 32t_q] \geq p^{32t_q/\ell}\cdot \Pr_{\cU}[T_q^i\leq 32t_q/\ell,T_q\leq 32t_q]\geq p^{32t_q/\ell}/2. 
\end{equation}
If Bob could \emph{detect} the event $\cW_q$, we would be 
done.
Indeed, let $C_2$ denote the set of (addresses and contents of) cells associated with all future epochs $j<i$, i.e., all the cells probed by $D$ succeeding epoch $i$.
Due to the geometrically decreasing sizes of epochs, sending $C_2$ requires less than $n_i/(w\cdot t_u)^{a-1}$ bits of communication. 
Since Bob has all the updates preceding epoch $i$, he can simulate the data structure and generate the correct memory state of $D$ right before epoch $i$. 
In particular, Bob knows for every cell, assuming it is not probed since epoch $i$ (thus associated with some epoch $j>i$), what its content will be.
Therefore, when he is further given the messages $(c_0,c_2)$, Bob would be able to simulate the data structure perfectly on query $q$, assuming the event $\cW_q$ occurs.
If Bob could detect $\cW_q$, he could simply output a random bit if it does not occur, and follow the data structure if it does.
This strategy would have already produced an advantage of $p^{32t_q/\ell}\geq \exp(-t_q\log^2 (w\cdot t_u)/\log n)$, which would have finished the proof. 
As explained in the introduction, Bob has no hope of certifying the occurrence of the event $\cW_q$,
hence we must take a fundamentally different approach for arguing that condition \eqref{eq_pr_all_black_sampled} can nevertheless be 
(implicitly) used  to devise a strategy for Bob with a nontrivial advantage.  This is the heart of the proof. 

To this end, note that, given a query $q\in \cQ'$, a received sample $c_0$ and all cells $c_2$ associated with some epoch $j<i$, Bob can simulate 
$D$ on his partial update sequence ($u_{-i}$), filling in the memory updates according to $c_0$ and $c_2$, and \emph{pretending} that all cells in 
the query-path of $q$ which are associated with epoch $i$ are actually sampled in $c_0$ (i.e., pretending that the event $\cW_q$ occurs). 
See Step 5 of Figure \ref{figure:pi} for the formal simulation argument. 
Let $M'(u_{-i}, c_0,c_2)$ denote the resulting memory state obtained by Bob's simulation in the figure, given $u_{-i}$ and his received sets of cells $c_0,c_2$. 

Now, let us consider the (deterministic) sequence of cells $S_q$ that $D$ would probe given query $q$ in the above simulation with respect to Bob's memory state $M'(u_{-i}, c_0,c_2)$.
Let us say that the triple $(u_{-i}, c_0,c_2)$ is \emph{good} for a query $q\in\cQ'$, if $\Pr_{\ep_i}[\cW_q|\ep_{-i}=u_{-i},C_0=c_0, C_2=c_2] \geq p^{32t_q/\ell}/4$ and $|S_q|\leq 32t_q$. 
That is, $(u_{-i},c_0,c_2)$ is good for $q$, if the posterior probability of $\cW_q$ is (relatively) high and $S_q$ is not too large.
By Equation~\eqref{eq_pr_all_black_sampled} and Markov's inequality, the probability that the triple $(u_{-i}, c_0,c_2)$ satisfies $\Pr_{\cU}[\cW_q, T_q\leq 32t_q|u_{-i},c_0,c_2]\geq p^{32t_q/\ell}/4$, is at least $p^{32t_q/\ell}/4$ 
(indeed, the expectation in \eqref{eq_pr_all_black_sampled} can be rewritten as $\E_{\ep_{-i},C_0,C_2}\Pr_{\ep_i}[\cW_q, T_q\leq 32t_q|\ep_{-i},C_0,C_2]$, since $C_2$ is a deterministic 
function of $\cU$). 
Note that when $\cW_q$ occurs, the value of $T_q$ is completely determined given $u_{-i}$, $c_0$ and $c_2$, in which case $|S_q|=T_q$, and 
thus the probability that $(u_{-i},c_0, c_2)$ is good is at least $p^{32t_q/\ell}/4$.
From now on, let us focus only on the case that $(u_{-i},c_0,c_2)$ that 
Alice sends is good, since Bob can identify whether $u_{-i},c_0,c_2$ is good based on $q$ and Alice's message, and if it is not, he will output a random bit. 

We caution that $S_q$ is simply a set of memory addresses in $M$, 
not necessarily the correct one -- in particular, while the \emph{addresses} of the cells $S_q$ are determined by the above simulation, 
the \emph{contents} of these cells (in $M$) are not --  they are a random variable of $\ep_{i}$, 
as the sample $c_0$ is very unlikely to contain all the associated cells).  
For any assignment $z \in [2^w]^{S_q}$ to the \emph{contents} of the cells in $S_q$, let us denote by 
\[\mu_q(z) := \Pr_{\ep_i}[S_q\leftarrow z | u_{-i},c_0,c_2]  \] 
the probability that the memory content of the sequence of cells $S_q$ is equal to $z$, conditioned on $u_{-i},c_0, c_2$. 

Every content assignment $Z=z$ to $S_q$, generates some posterior distribution on the \emph{correct} 
query path (i.e., with respect to the \emph{true} memory state $M$) and therefore on the output $\cP(q,\cU)$ of the query $q$ with 
respect to $\cU$. Hence we may look at the joint probability distribution of the event  ``$\cP(q,\cU)=1$" and the assignment $Z$ 
which is 
\[\eta_q(z) := \Pr_{\ep_i}[\cP(q,\cU) = 1 , \; S_q\leftarrow z \mid  u_{-i},c_0,c_2 ]. \] 
Now, consider the function 
\begin{align}\label{eq_def_f}
f(z) = f^q_{u_{-i},c_0,c_2}(z) &:= \eta_q(z)   -  \frac{1}{2}\cdot \mu_q(z).
\end{align} 
Equivalently, conditioned on $u_{-i}$, $c_0$ and $c_2$, $f(z)$ is the bias of the random varaible $\cP(q,\cU)$ conditioned on $S_q\leftarrow z$, multiplied by the probability of $S_q\leftarrow z$.

Note that, since $\eta_q(z) \leq \mu_q(z)$ for every assignment $z$, we have $|f(z)|\leq \mu_q(z)/2$, 
and since $\mu_q(z)$ is a probability distribution, this fact implies that: (i) $\sum_{z} |f(z)|\leq \frac{1}{2}$. Furthermore, we shall argue 
that $\Pr[\cW_q\mid u_{-i},c_0,c_2]\geq p^{32t_q/\ell}/4$ (as we always condition on good $u_{-i},c_0,c_2$), in which case 
the contents of $S_q$ are completely determined by $u_{-i},c_0,c_2$ 
(we postpone the formal argument to the Analysis section below). 
Denoting by $z^*$ the content assignment to $S_q$ induced by $u_{-i},c_0,c_2$, we observe that 
conditioned on $\cW_q$,  $S_q$ will be precisely the correct set of cells probed by $D$ on $q$, in which case $\cP(q,\cU)$ is determined by $z^*,q,u_{-i},c_0,c_2$. 
Formally, this fact means that: (ii) $|f(z^*)|=\frac{1}{2}\cdot \Pr[S_q\leftarrow z^*\mid u_{-i},c_0,c_2]\geq \Omega(p^{32t_q/\ell})$. 

Conditions (i)$+$(ii) above imply that $f = f^q_{u_{-i},c_0,c_2}$ satisfies the premise of the Peak-to-Average Lemma (Lemma~\ref{lem_peak_to_avg})  
with $\Sigma := [2^w], k:= O(t_q), \eps := \Omega(p^{32t_q/\ell})=\exp(-t_q \log^2 (w\cdot t_u)/\log n)$. Recall that the lemma guarantees there is a 
not-too-large subset $Y\subseteq S_q$ 
of coordinates ($=$ addresses) of $S_q$, which Bob can \emph{privately compute},\footnote{Indeed, $Y$ is only a function of $q$, $f^q_{u_{-i},c_0,c_2}$, 
$c_0$,$c_2$ and the prior distribution on $\cU$, and Bob possesses all this information.}  
such that if the values of the coordinates in $Y$ are also revealed, then the \emph{conditional} expectation of $f^q_{u_{-i},c_0,c_2}$ 
$\left(\text{namely, }  \E_{S_q|_Y} \left| \Pr_{\ep_i}  \left[ \cP(q,\cU)=1 \mid u_{-i},c_0,c_2,S_q|_Y \right]  \; - \; 1/2 \right|\right)$, which is the average of 
Bob's ``maximum-likelihood" estimate for $\cP(q,\cU)$, is non-negligible (the formal details are postponed to the Analysis section below). 

Given this insight, a natural strategy for the players is for Alice to further send Bob the contents of cells in the subset $Y$. 
While Alice does not know the subset $Y$,\footnote{Indeed, $Y$ is a function of $q$.} 
she can use \emph{public randomness} to sample yet another random set $C_1$ of cells from the \emph{entire} memory $M$, 
where now \emph{every} cell is sampled with equal probability $p$, 
and send the subset of $C_1$ that is associated with epoch $i$ to Bob. (Note that it is important that this time the players use public randomness to subsample from the entire memory 
state $M$, since Alice does not know $Y$ and yet Bob must be absolutely certain that all cells in $Y$ were subsampled. Notwithstanding, 
to keep communication low, it is crucial that Alice sends Bob only the
contents of cells associated with epoch $i$).
Since $|Y|$ is guaranteed to be relatively small (of order $O(\sqrt{k\log(1/\eps)})$), 
the probability $p^{|Y|}$ that all cells  in $Y$ get sampled will be sufficiently noticeable, in which case we shall argue that Bob's 
maximum-likelihood strategy will output the correct answer $\in\{0,1\}$ with the desired nontrivial advantage.  
The formal one-way protocol $\pi$ that the parties execute is described in Figure \ref{figure:pi}.


\begin{figure}
\begin{tabular}{|l|}
\hline
\begin{minipage}{\algwidth}
\vspace{1ex}
\begin{center}
\textbf{One-way protocol $\pi$ for $G^i_\cP$}
\end{center}
\vspace{0.5ex}
\end{minipage}\\
\hline
\begin{minipage}{\algwidth}
\vspace{1ex}

Henceforth, by ``sending a cell'', we mean sending the address and (up to date) content of the cell in $M$.
\paragraph*{Encoding:}
\begin{enumerate}
    \item Alice generates the memory state $M$ of $D$ by simulating the data structure on $\cU$, and computes the associated epoch for each cell.
    \item Alice samples each cell associated with epoch $i$ independently with probability $p$. 
    Let $c_0$ be the set of sampled cells.
    If $|c_0|>2p|\ep_i|\cdot t_u$, Alice sends a bit $0$ and aborts.
    Otherwise, she sends a bit $1$, followed by all cells in $c_0$. 
    \item Alice uses \emph{public randomness} to sample \emph{every cell in $M$} independently with probability $p$. 
    Let $c_1$ be the set of sampled cells.
    If there are more than $2p|\ep_i|\cdot t_u$ cells in $c_1$ that are associated with epoch $i$,
    Alice sends a bit $0$ and aborts.
    Otherwise, she sends a bit $1$, followed by all cells in $c_1$ that are \emph{associated with epoch $i$}.
    \item Alice sends Bob all cells associated with epoch $j$ for all $j<i$, i.e., all the cells probed by $D$ succeeding epoch $i$.
    Denote this set of (address and contents of) cells by $c_2$.
\end{enumerate}

\paragraph*{Decoding:}

\begin{enumerate}
    \setcounter{enumi}{4}
    \item Given his query $q\in \cQ$, Bob simulates the data structure $D$ on $u_{>i}$ and obtains a memory state $M_0$. 
    He updates the contents of $c_0$ and $c_2$ in $M_0$, obtains a memory state $M'=M'(u_{-i}, c_0,c_2)$,  
    and then simulates the query algorithm of $D$ on query $q$ and memory state $M'$. Let $S_q$ be the set of (memory addresses of) cells probed 
    by $D$ 
    in this simulation. If any of the following events occur, Bob outputs a random bit and aborts:
    \begin{enumerate}[(i)]
        \item $q \notin \cQ'$, 
        \item Bob receives a bit $0$ before $c_0$ or $c_1$,
        \item $(u_{-i},c_0,c_2)$ is not good for $q$.
    \end{enumerate}
    \item 
    Let $Y \subset S_q$ be a subset of cells of size $ \kappa := |Y| \leq O\left(\sqrt{k\cdot \log 1/\eps}\right)$ 
    guaranteed by Lemma~\ref{lem_peak_to_avg}, when applied with $f := f^q_{u_{-i},c_0,c_2} $, $\Sigma:=[2^w]$, $k:= |S_q|\leq 32t_q, \eps := p^{32t_q/\ell}/4$.  
    
    (recall that Bob can privately compute the set $Y$). 
    \item If $Y\nsubseteq c_1$ (i.e., if the sample $c_1$ sent by Alice does \emph{not} contain all cells in $Y$), 
    Bob outputs a random bit. 
    Otherwise, let $y \in [2^w]^{Y}$ denote the content  of the cells $Y$ according to $c_1$.  
    Let $S_q|_Y\leftarrow y$ denote the event that the memory content of $Y$ is assigned the value $y$.
    Bob outputs $1$ iff  
    \[ 
    \Pr_{\ep_i}\left[\cP(q,\cU) = 1 \mid u_{-i},c_0,c_2, S_q|_Y\leftarrow y \right]  
    \;\; > \;\; 1/2 . \] 
    Otherwise, Bob outputs $0$.
\end{enumerate}
\vspace{0.3ex}
\end{minipage}\\
\hline
\end{tabular}
\caption{The one-way weak simulation protocol of data structure $D$. }\label{figure:pi}
\end{figure}


\paragraph{Analysis.} We now turn to the formal analysis of the protocol $\pi$. We need to show     
\begin{itemize}
\item (Communication cost)  $\|\pi\| \leq O(n_i/(w\cdot t_u)^{a-1})$ . 
\item (Correctness) 
$\Pr_{G^i_\cP \sim \cU, q\in_R\cQ} \left[\pi(G^i_\cP) = \cP(q,\cU)\right] \geq 1/2 + \exp\left(-t_q \log^2 (w\cdot t_u)/\sqrt{\log n}\right)$.
\end{itemize}

\paragraph{Communication.}
In both Step 2 and Step 3, Alice sends at most $2p|\ep_i| t_u\cdot (2w)+1$ bits. In Step 4, 
Alice sends at most $|\ep_{<i}|\cdot t_u\cdot (2w) = \sum_{j < i} |\ep_j| \cdot t_u\cdot (2w) $ bits. 
Since $|\ep_j|=n_j=\beta^j$, the total communication cost is at most
\[
	O(p\cdot n_i\cdot t_uw)+O(\beta^{i-1}\cdot t_uw)\leq O(n_i/(w\cdot t_u)^{a-1}).
\]

\paragraph{Correctness.}{
Let $\pi'$ be the variant of the protocol $\pi$ in which, when executing Step 2 and Step 3, Alice ignores the condition of 
whether the samples $C_0$ or $C_1$ exceed the specified size limit, 
i.e., she always sends a bit 1 followed by all sampled cells. For simplicity of analysis, 
we will first show that $\pi'$ has the claimed success probability, and then show that the impact of the above event (i.e., conditioning on $C_0$ 
and $C_1$ being within the size bound) is negligible, as it occurs with extremely high probability.   

We first claim that the probability (over $\cU$ and an average query $q\in_R Q$) that $\pi'$ reaches Step $6$ is 
not too small.
By \eqref{eq_qury_time_epoch_i} and Markov's inequality, and by the discussion below \eqref{eq_pr_all_black_sampled},  
the probability that $q\in \cQ'$ and $(u_{-i},c_0,c_2)$ is ``good" for $q$ is at least $\Omega(p^{32t_q/\ell})\geq \exp(-t_q \log^2 (w\cdot t_u)/\log n)$.
This is precisely the probability that $\pi'$ reaches Step 6.

We now calculate the success probability of $\pi'$ conditioned on reaching Step 6.
To this end, fix a set $Y\subseteq S_q$ of size $\kappa$. 
Then by Step $7$,  the success probability of $\pi'$ conditioned on $u_{-i},c_0,c_2$ and the event  ``$Y\subseteq C_1$" is
\begin{align} \label{eq_advantage}
	\frac{1}{2}\; &+ \E_{S_q|_Y} \left| \Pr_{\ep_i}  \left[ \cP(q,\cU)=1 \mid u_{-i},c_0,c_2,S_q|_Y \right]  \; - \; 1/2 \right| \nonumber \\
	=\frac{1}{2}\; &+\sum_{y\in [2^w]^Y} \Pr_{\ep_i}[(S_q|_Y\leftarrow y)\mid u_{-i},c_0,c_2]\cdot \left| \Pr_{\ep_i}  
	\left[ \cP(q,\cU)=1 \mid (S_q|_Y\leftarrow y), u_{-i},c_0,c_2 \right]  \; - \; 1/2 \right| \nonumber \\
	=\frac{1}{2}\; &+\sum_{y\in [2^w]^Y} \left| \Pr_{\ep_i}  \left[ \cP(q,\cU)=1,(S_q|_Y\leftarrow y) \mid u_{-i},c_0,c_2 \right]  
	\; - \; \frac{1}{2}\cdot \Pr_{\ep_i}[(S_q|_Y\leftarrow y)\mid u_{-i},c_0,c_2] \right|  \nonumber \\
	=\frac{1}{2}\; &+\sum_{y\in [2^w]^Y} \left| \sum_{z\in [2^w]^{S_q} \; : \; z|_Y=y} \left(\Pr_{\ep_i}\left[ \cP(q,\cU)=1,(S_q\leftarrow z) \mid u_{-i},c_0,c_2 \right]  \; - \; \frac{1}{2}\cdot \Pr_{\ep_i}[S_q\leftarrow z\mid u_{-i},c_0,c_2]\right) \right|  \nonumber \\ 
		=\frac{1}{2}\; &+\sum_{y\in [2^w]^Y} \left| \sum_{z\in [2^w]^{S_q} \; : \; z|_Y=y} f^q_{u_{-i},c_0,c_2}(z) \right|  \; , 
\end{align}
where the last transition is by the definition of $f^q_{u_{-i},c_0,c_2}$ in \eqref{eq_def_f}. 
Note that for any $z$, it holds that $|f(z)|\leq \frac{1}{2}\cdot \Pr[S_q\leftarrow z\mid u_{-i},c_0,c_2]$. 
Thus, $\sum_{z\in [2^w]^{S_q}} |f(z)|\leq \frac{1}{2}$. 
On the other hand, since we always condition on good $(u_{-i},c_0,c_2)$, we have $\Pr[\cW_q\mid u_{-i},c_0,c_2]\geq p^{32t_q/\ell}/4$. 
That is, with probability at least $p^{32t_q/\ell}/4$ all cells in $S_q$ associated with epoch $i$ are contained in $c_0$.
In this case, the contents of $S_q$ are completely determined by $u_{-i},c_0,c_2$. 
Indeed, the contents of the cells associated with epoch $<i$ are determined by $c_2$;
the cells associated with epoch $i$ are determined by $c_0$;
the remaining cells are determined by $u_{>i}$.
Let $z^*$ denote the assignment to $S_q$, induced by $u_{-i}$ and the contents of $c_0,c_2$ conditioned on the occurrence of $\cW_q$.
By the definition of $S_q$, when $\cW_q$ happens, $S_q$ will be exactly the set of cells the data structure probes.
Thus, the output of $q$ is also determined.
We therefore have $|f(z^*)|=\frac{1}{2}\cdot \Pr[S_q\leftarrow z^*\mid u_{-i},c_0,c_2]\geq \Omega(p^{32t_q/\ell})$.
We conclude that the function $f=f^q_{u_{-i},c_0,c_2}$ satisfies the premise of the Peak-to-Average lemma (Lemma~\ref{lem_peak_to_avg}) with
\begin{itemize}
	\item
		$\Sigma=[2^w]$;
	\item
		$k=|S_q|\leq O(t_q)$;
	\item
		$\eps=p^{32t_q/\ell}/4\geq\exp(-t_q\log^2(w\cdot t_u)/\log n)$.\footnote{We used the fact that $\ell = \Theta(\log_{\beta} n)$ and $\beta = (w\cdot t_u)^{\Theta(1)}$.}
\end{itemize}
Without loss of generality, we may assume $\log(w\cdot t_u)\ll \sqrt{\log n}$, and thus $\eps\in[2^{-O(k)},1]$.\footnote{In fact, if $\log (w\cdot t_u)\geq \Omega(\sqrt{\lg n})$, 
the right-hand side of the inequality in the theorem statement is less than $p^{t_q}$, hence the statement becomes trivial. Indeed, with probability $p^{t_q}$, Alice 
samples \emph{all} cells probed by the data structure on query $q$.}
Therefore, the lemma guarantees there is a set $Y \subset S_q$ of cells that 
has size at most 
\[ |Y|=\kappa \leq O\left(\sqrt{k\log 1/\eps}\right)
\leq O\left(t_q\log (w\cdot t_u)/\sqrt{\log n}\right) \; , \]
for which 
\[
\sum_{y\in [2^w]^Y} \left| \sum_{z\in [2^w]^{S_q} \; : \; z|_Y=y} f^q_{u_{-i},c_0,c_2}(z) \right| \geq \exp\left(-t_q\log (w\cdot t_u)/\sqrt{\log n}\right) \; .
\]
This justifies Step 6 of the protocol. 
It follows that, for any $q\in \cQ'$, the probability that the sample 
$C_1$ of cells contains the set $Y$ is at least 
\begin{equation}\label{eq_pr_abort_step_7}
\Pr_{C_1}\left[Y \subseteq C_1\right] = p^{|Y|} = p^{O\left(t_q\log (w\cdot t_u)/\sqrt{\log n}\right)} = \exp\left(-t_q\log^2 (w\cdot t_u)/\sqrt{\log n}\right). 
\end{equation}
Equation \eqref{eq_advantage} therefore implies that, conditioned on the event that $|Y| \subseteq C_1$, 
the probability that $\pi'$ outputs a correct answer is 
\[
1/2+\exp\left(-t_q\log (w\cdot t_u)/\sqrt{\log n}\right) , 
\]
and combining this with \eqref{eq_pr_abort_step_7} and the probability that $\pi'$ reaches Step 6, we conclude that the overall success probability of $\pi$, 
conditioned on the protocol not aborting when $c_0$ or $c_1$ is too large, is 
\begin{equation}\label{eq_adv_pi_overall_conditioned_not_abort}
1/2+\exp\left(-t_q\log^2 (w\cdot t_u)/\sqrt{\log n}\right) . 
\end{equation}

To finish the proof, it therefore suffices to argue that the probability that $\pi$ aborts due to this event is tiny. 
To this end, let $A_i$ denote the random variable representing the number of associated cells with epoch $i$.  
We know that $A_i \leq |\ep_i|\cdot t_u = n_i\cdot t_u$ (since the worst-case update time of $D$ is $t_u$ by assumption). 
Now, let $\cE_0$ denote the event that Alice's sample in Step 2 of the protocol is too large, i.e., that $``|C_0| > 2p|\ep_i|\cdot t_u"$. 
Similarly, let $\cE_1$ denote the event that in Step 3 of the protocol, $``|C_1| > 2p|\ep_i|\cdot t_u"$.
Denote $\cE :=  \cE_0  \vee \cE_1$ (note that this is the event (ii) in Step 5 of $\pi$).
Since both sets $C_0$ and $C_1$ are \emph{i.i.d} samples where each cell is sampled independently with probability $p$, a standard Chernoff bound 
implies that 
\begin{equation}\label{eq_c_0_c_1_chernoff}
\Pr[\cE] \leq 2\Pr\left[|C_0| \geq 2\E\left[|C_0|\right]\right] \leq \exp(-p(n_i\cdot t_u))\leq \exp(-n_i/(w\cdot t_u)^{a}). 
\end{equation}
Finally, since $i\geq \ell/2$ and thus $n_i\geq n^{1/2}\geq t_q (w\cdot t_u)^{a+1}$, by~\eqref{eq_adv_pi_overall_conditioned_not_abort}, \eqref{eq_c_0_c_1_chernoff} and a union bound, we conclude that 
\[
\begin{aligned}
	\Pr_{\cU,q}[\pi(q) \neq \cP(q,\cU)]&\leq 1/2-\exp\left(-t_q\log^2 (w\cdot t_u)/\sqrt{\log n}\right)+\Pr[\cE]\\
	&\leq 1/2-\exp\left(-t_q\log^2 (w\cdot t_u)/\sqrt{\log n}\right)+\exp(-t_q\cdot (w\cdot t_u))\\
	&\leq 1/2-\exp\left(-t_q\log^2 (w\cdot t_u)/\sqrt{\log n}\right) , 
\end{aligned}
\]
which completes the proof of the entire theorem.}

\end{proof}

While Theorem~\ref{thm_weak_simulation_informal} is very clean, we shall need a slightly more
technical version of it for some of our lower bound proofs. The
following corollary follows directly by examining the proof of
Theorem~\ref{thm_weak_simulation_informal}:

\begin{theorem}[One-Way Weak Simulation of Epoch $i$]	\label{thm_weak_simulation_epoch}
Let $\cP$ be a dynamic boolean data structure problem, with $n$ random updates grouped into epochs $\cU = \{\ep_i\}_{i=1}^\ell$, such that $|\ep_i|=\beta^i$
followed by a single query $q\in \cQ$. If $\cP$ admits a dynamic data
structure $D$ with worst-case update time $t_u$ and average (over
$\cQ$) expected query time $t_q$, such that for some
epoch $i \in [\ell/2,\ell]$
it holds that
$$
\frac{1}{|\cQ|} \sum_{q \in Q} \E[T^i_q] \leq 2t_q/\ell,
$$
then if $t_q(w\cdot t_u)^{a+1} \leq n_i$, we have
\[ \suc\left(G^i_\cP, \;\cU, \; n_i/(w\cdot t_u)^{a-1}\right) \geq  \exp\left(-t_q \log^2 (w\cdot t_u)/\sqrt{\log n}\right)\]
as long as $\beta = (w \cdot t_u)^{\Theta(1)} \geq (w\cdot t_u)^a$ for a constant $a>1$.
\end{theorem}



\subsection{Proof of the Peak-to-Average Lemma} \label{subsec_PTA_lemma}

In this subsection we prove our key technical lemma, which is required to complete the proof of Theorem \ref{thm_weak_simulation_informal}. 

\begin{restate}[Lemma~\ref{lem_peak_to_avg}]
Let $f : \Sigma^k \to \R $ be any real function on the length-$k$ strings over alphabet $\Sigma$, satisfying: 
\begin{enumerate}[(i)]
\item  $\sum_{z\in\Sigma^k}|f(z)|\leq 1$; and 
\item $\max_{z\in\Sigma^k}|f(z)| \geq \eps$
\end{enumerate}
for some $\eps\in [2^{-O(k)}, 1]$.
Then there exists a subset $Y$ of indices, $|Y| \leq O\left(\sqrt{k\cdot \log 1/\eps}\right)$, such that 
\[
    \sum_{y\in\Sigma^Y}\left|\sum_{z|_Y=y}f(z)\right|\geq\exp\left(-\sqrt{k\cdot \log 1/\eps}\right). 
\]

\end{restate} 

\vspace{0.1in}

The lemma is tight, as shown in Section \ref{app_pf_tight_PtA} of the Appendix. 
We first provide a high-level overview of the proof of Lemma \ref{lem_peak_to_avg}, and then proceed 
to the formal proof. 
The first observation is that we may without loss of generality assume that $\Sigma=\{0,1\}$ (intuitively, the larger the alphabet is, the more 
information we will learn upon revealing the values of $Y$). 
Assuming $f$ is defined on a boolean hypercube, the high-level intuition for the proof is that we can \emph{multiply} $f(z)$ 
by a low-degree polynomial $Q(z)$ that point-wise approximates the $\AND$ function to within an additive error $\eps$, 
a.k.a.,  a (variant of the) discrete Chebychev polynomial, which 
has the effect of preserving the $\ell_\infty$ value of $f$ but 
exponentially ``dampening" the magnitude of all remaining values,  
thereby making the maximum (constant) value of $(f\cdot Q)(z)$ dominate the \emph{sum} of all remaining values. Since the degree of $Q(z)$ (required to ensure 
the latter property) is $d := O(\sqrt{k\log(1/\eps)})$, $Q$ itself can be written as the sum of 
at most ${k\choose d}= \exp(\tilde{O}(\sqrt{k\cdot \log 1/\eps}))$ monomials, hence 
\emph{one of these monomials} (which can be viewed as some \emph{specific subset} of $\leq d = O(\sqrt{k\log(1/\eps)})$ coordinates) must 
 account for at least $\gtrsim \exp(-\sqrt{k\cdot \log 1/\eps})$ fraction of the total sum $\sum_z f(z)\cdot Q(z)$, 
so \emph{fixing} this particular monomial's coordinates (which is the small subset $Y$ we are looking for) to $1$ must contribute the aforementioned 
quantity to the average \emph{of $f$}.  

\subsubsection{The formal proof} \label{subsubsec_formal_PTA}

As discussed above, a central ingredient of the proof is the existence of low-degree polynomials with ``threshold" phenomena, 
commonly known as \emph{Chebychev polynomials}. In particular, the following lemma states that there is a low-degree multivariate 
polynomial that \emph{point-wise} approximates the \emph{AND} function on the $k$-dimensional hypercube to within small error (i.e., in the 
$\ell_\infty$ sense).
The following lemma, which is translating the quantum algorithm in~\cite{BCWZ99}, asserts the existence of such polynomials.

\begin{lemma}\label{lem:discrete_chebyshev}
For any $k$ and $M$ satisfying $2\leq M\leq 2^{O(k)}$, there exists a polynomial $Q=Q_{k,M}(x_1,\ldots,x_k)$ such that
\begin{enumerate}[(i)]
    \item
        $Q$ has total degree $O(\sqrt{k\log M})$;
    \item
        $|Q(0^k)|\geq M$;
    \item
        $\forall x\in\{0,1\}^k\setminus\{0^k\},\left|Q(x)\right|\leq 1$;
    \item
        The sum of absolute values of all coefficients is at most $\exp(\sqrt{k\log M})$.
\end{enumerate}
\end{lemma}



\vspace{0.1in}
The proof of the lemma can be found in Appendix~\ref{app_pf_cheby}. 
We are now ready to prove the Peak-to-Average Lemma.  

\begin{proof}[Proof of Lemma~\ref{lem_peak_to_avg}]

We first show that without loss of generality, we may assume that $\Sigma=\{0,1\}$ and $|f(0)|\geq \eps$.
In general, let $z^*\in\Sigma^k$ be any point with large absolute $f$-value: $|f(z^*)|\geq \eps$.
Define $h:\{0,1\}^k\to \R$ as follows:
\[
h(x)=\sum_{z\in\Sigma^k:z_i=z^*_i \mathrm{iff} x_i=0} f(z).
\]
It is easy to verify that $$\sum_{x\in\{0,1\}^k}|h(x)|\leq \sum_{z\in\Sigma^k} |f(z)|\leq 1$$ and $|h(0)|=|f(z^*)|\geq \eps$, i.e., $h$ satisfies both conditions in the lemma statement.
Moreover, for any subset $Y$ of indices, we have
\[
\begin{aligned}
\sum_{y\in\{0,1\}^Y}\left|\sum_{x|_Y=y}h(x)\right|&\leq\sum_{y\in\{0,1\}^Y}\left|\sum_{x|_Y=y,z\in\Sigma^k:z_i=z^*_i \mathrm{iff} x_i=0} f(z)\right| \\
&\leq \sum_{y'\in\Sigma^Y}\left|\sum_{z|_Y=y'} f(z)\right|.
\end{aligned}
\]

Thus, it suffices to prove the lemma assuming $\Sigma=\{0,1\}$ and $|f(0)|\geq \eps$. 

\medskip

Let $Q=Q_{k,2/\eps}$ be a polynomial with all four properties in Lemma~\ref{lem:discrete_chebyshev} with $M=2/\eps$.
Since $2^{-O(k)}\leq \eps\leq 1$, such polynomial exists and has degree $d\leq O(\sqrt{k\log M})$.
Without loss of generality, we may assume $Q$ is multi-linear.\footnote{$Q$ is only evaluated on $\{0,1\}^k$, and all four properties are preserved when replacing $x_i^2$ by $x_i$.}
Thus, let
\[
Q(x)=\sum_{Y\subseteq [k],|Y|\leq d}\alpha_Y\cdot\prod_{i\in Y} x_i.
\]
By Property (iv), we have 
\begin{equation}\label{eq:p4}
\sum_Y |\alpha_Y|\leq \exp(\sqrt{k\log M}).
\end{equation}

Now, consider the function 
\[ g(x) := f(x)\cdot Q(x) . \] 
By the premise of the lemma, we have that 
$|g(0^k)| \geq \frac{2}{\eps}\cdot \eps = 2$, but 
\[ \left|\sum_{x\in \{0,1\}^k\setminus \{0^k\}} g(x)\right| \leq \sum_{x\in \{0,1\}^k\setminus \{0^k\}}\left| g(x)\right| \leq 
\sum_{x\in \{0,1\}^k\setminus \{0^k\}}\left| f(x)\right| \leq 1 , \]  
by the triangle inequality and $|Q(x)|\leq 1$ for $x\neq 0^k$. Therefore,  we have 
\begin{align}\label{eq_sum_val} 
\left|\sum_{x\in \{0,1\}^k} g(x)\right| \geq 2 - 1 = 1. 
\end{align}

On the other hand, we have
\begin{align*}
\sum_{x\in \{0,1\}^k} g(x) &= \sum_{x\in \{0,1\}^k} f(x) \cdot Q(x)  \\
&= \sum_{x\in \{0,1\}^k} f(x) \cdot \left(\sum_{Y\subseteq [k],|Y|\leq d} \alpha_Y\cdot \prod_{i\in Y}x_i\right)  \\
& =  \sum_{Y\subseteq [k],|Y|\leq d} \alpha_Y \cdot  \left(    \sum_{x\in \{0,1\}^k} f(x)\cdot \prod_{i\in Y}x_i \right)   \\ 
& =  \sum_{Y\subseteq [k],|Y|\leq d} \alpha_Y\cdot  \sum_{x\in \{0,1\}^k : x_i = 1\textrm{ for }i\in Y} f(x).
\end{align*}
By Equation~\eqref{eq_sum_val} and Equation~\eqref{eq:p4}, there must exist some $Y\subseteq [k]$ and $|Y|\leq d$
for which  
\[ \left|\sum_{x\in \{0,1\}^k : x_i= 1\textrm{ for }i\in Y} f(x)\right| \geq \exp(-\sqrt{k\log 1/\eps}). \] 
Thus, we have
\[ \sum_{y\in\Sigma^Y}\left|\sum_{x\in\{0,1\}^k : x|_Y = y} f(x)\right| \geq  \exp\left(-\sqrt{k\cdot \log 1/\eps}\right) , \] 
and $|Y| \leq d = O(\sqrt{k\cdot \log 1/\eps})$, as claimed. 
\end{proof}

\ifx\mainfile\undefined
\bibliographystyle{alpha}
\bibliography{refs}
\end{document}
\fi

\section{Boolean Polynomial Evaluation}
\label{sec:polyeval}
In this section, we prove our first concrete lower bound using our new technique. Let $\cP$ be the dynamic least-bit polynomial evaluation problem over the Galois field $\GF(2^d)$ (as defined in Section~\ref{sec:concreteproblems}). Recall that the data structure problem $\cP$ is defined as follows: A degree $n \leq 2^{d/4}$ polynomial $P(x) = \sum_{i=0}^n a_i x^i$ over $\GF(2^d)$ is initialized with all $n+1$ coefficients $a_i$ being $0$. An update is specified by a tuple $(i, b)$ where $i \in [n+1]$ is an index and $b$ is an element in $\GF(2^d)$. It changes the coefficient $a_i$ such that $a_i \gets a_i + b$ (where addition is over $\GF(2^d)$). A query is specified by an element $y \in \GF(2^d)$ and one must return the least significant bit of $P(y)$. Note that we make no assumptions on the concrete representation of the elements in $\GF(2^d)$, only that the elements are in a bijection with $\{0,1\}^d$ so that precisely half of all elements in $\GF(2^d)$ have a $0$ as the least significant bit.

We consider the following random sequence of updates $\cU = (\ep_\ell,\ep_{\ell-1},\ldots,\ep_1)$, where $|\ep_i|=n_i = \beta^i$ for some $\beta = (wt_u)^{\Theta(1)}$. The maintained polynomial $P(x)$ has degree $n = \beta^\ell-1$, and thus the number of updates is $\Theta(n)$. The $n_i$ updates $\ep_i$ are $(0,b_0),\dots,(n_{i-1},b_{n_i-1})$ in that order, where the $b_j$'s are chosen independently and uniformly at random from $\GF(2^d)$. The query $q$ is chosen as a uniform random element of $\GF(2^d)$.

Invoking Theorem~\ref{thm_weak_simulation_informal}, the existence of a a dynamic data structure for $\cP$ with worst case update time $t_u$ and 
expected query time $t_q$ under $\cU$ implies that either $t_q(w t_u)^{\Theta(1)} > n^{1/2}$ or for some $i \in \{\ell/2,\dots,\ell\}$, we have
\[ \suc\left(G^i_\cP, \;\cU, \; n_i/(w\cdot t_u)^{\Theta(1)}\right) \geq  \exp\left(-t_q \log^2 (w\cdot t_u)/\sqrt{\log n}\right) . \]
In the first case, we are already done as we have a lower bound of $t_q = \Omega(\sqrt{n}/(w t_u)^{O(1)})$.
We thus set out to prove an upper bound on $\suc\left(G^i_\cP, \;\cU, \; n_i/(w\cdot t_u)^{\Theta(1)}\right)$ for any epoch $i \in \{\ell/2,\dots,\ell\}$ (assuming $t_q(w t_u)^{\Theta(1)} \leq n^{1/2}$).

\begin{lemma}
\label{lem:noAdv}
For any epoch $i \in \{\ell/2,\dots,\ell\}$, we have $\suc\left(G^i_\cP, \;\cU, \; o(n_i d) \right) \leq 2^{-d/8}$.
\end{lemma}

Before proving the lemma, let us use it to derive our lower bound. We see that it must be the case that:
\begin{eqnarray*}
\exp\left(-t_q \log^2 (w\cdot t_u)/\sqrt{\log n}\right) \leq 2^{-d/8} \Rightarrow \\
t_q \log^2 (w\cdot t_u)/\sqrt{\log n} = \Omega(d) \Rightarrow \\
t_q = \Omega(d \sqrt{\lg n}/\log^2 (w\cdot t_u)).
\end{eqnarray*}
Thus Theorem~\ref{thm:polyeval} follows. Proving Lemma~\ref{lem:noAdv} is the focus of Section~\ref{sec:noAdv}.

\subsection{Low Advantage on Epochs}
\label{sec:noAdv}
Let $i \in \{\ell/2,\dots,\ell\}$. Recall that in the communication game $G^i_\cP$ for the random update sequence $\cU$, Alice receives all updates of all epochs and Bob receives all updates of all epochs except epoch  $i$. Bob also receives the query $q$. Let $\pi$ be a one-way randomized protocol in which Alice sends $o(n_i d)$ 
bits to Bob, and suppose that $\pi$ achieves an advantage of $\eps$ w.r.t $\cU$ and $q$. Since the query $q$ and the updates $\ep_i$ are independent of the updates of epochs $\ep_\ell,\dots,\ep_{i+1},\ep_{i-1},\dots,\ep_1$, we can fix the random coins of the protocol and fix the updates of all epochs except epoch $i$, such that for the resulting deterministic protocol $\pi^*$ and fixed update sequence $u_\ell,\dots,u_{i+1},u_{i-1},\dots,u_1$ we have that Alice never sends more than $o(n_id)$ bits and $\Pr_{\ep_i,q}[v_{\pi^*} = P_0(q)] \geq 1/2+\eps$. Here $P_0$ is the least significant bit of $P(q)$, where $P$ is the polynomial resulting from performing the updates $u_\ell,\dots,u_{i+1},\ep_i,u_{i-1},\dots,u_1$. Recall that the random variable $v_{\pi^*}$ is Bob's output when running the deterministic protocol $\pi^*$ on $u_\ell,\dots,u_{i+1},\ep_i,u_{i-1},\dots,u_1$ and $q$.

Let $M_{\pi^*}(\ep_i)$ denote the message sent by Alice in procotol $\pi^*$ on updates $u_\ell,\dots,u_{i+1},\ep_i,u_{i-1},\dots,u_1$. Then $v_{\pi^*} = v_{\pi^*}(M_{\pi^*}(\ep_i),q)$ is determined from $M_{\pi^*}(\ep_i)$ and $q$ alone (since the updates of other epochs are fixed). For each of the possible messages $m$ of Alice, define the vector $\chi_m$ having one coordinate per $x \in \GF(2^d)$. The coordinate $\chi_m(x)$ corresponding to some $x$ has the value $-1$ if $v_{\pi^*}(m,x)=0$ and it has the value $1$ otherwise. Similarly define for each sequence of updates $u_i \in \supp(\ep_i)$ the vector $\psi_{u_i}$ having one coordinate $\psi_{u_i}(x)$ per $x \in \GF(2^d)$, where the coordinate corresponding to some $x$ takes the value $-1$ if the correct answer to the query $x$ is $0$ after the update sequence $u_\ell,\dots,u_{i+1},u_i,u_{i-1},\dots,u_1$ and taking the value $1$ otherwise. Since $\pi^*$ has advantage $\eps$ and $q$ is uniform over $\GF(2^d)$, we must have
$$
\E_{\ep_i}[\langle \psi_{\ep_i} , \chi_{M_{\pi^*}(\ep_i)} \rangle ] = ((1/2+\eps) - (1/2-\eps))2^d = \eps 2^{d+1}.
$$
This in particular implies that if we take the absolute value of the inner product, we have
$$
\E_{\ep_i}[|\langle \psi_{\ep_i} , \chi_{M_{\pi^*}(\ep_i)} \rangle | ] \geq \eps 2^{d+1}.
$$
As we will show later, this implies the following:
\begin{lemma}\label{lem:findingM}
There is some $m \in \supp(M_{\pi^*}(\ep_i))$ such that we have both
\begin{itemize}
\item $\E_{\ep_i}[|\langle \psi_{\ep_i} , \chi_{M_{\pi^*}(\ep_i)} \rangle | \mid M_{\pi^*}(\ep_i) = m] \geq \eps 2^d.$
\item $\Pr_{\ep_i}[M_{\pi^*}(\ep_i) = m] \geq |\supp(M_{\pi^*}(\ep_i))|^{-1}\eps/2.$
\end{itemize}
\end{lemma}

Consider such an $m$ and the corresponding vector $\chi_m$. We examine the following $k$'th moment for an even $k$ to be determined:
\begin{eqnarray*}
\E_{\ep_i}[ \langle \psi_{\ep_i}, \chi_m \rangle^k] &=& \\
\sum_{S \in \GF(2^w)^k} \E_{\ep_i} \left[\prod_{x \in S} \psi_{\ep_i}(x) \chi_m(x)\right] &=& \\
\sum_{S \in \GF(2^w)^k} \E_{\ep_i} \left[\prod_{x \in S} \psi_{\ep_i}(x) \right] \prod_{x \in S} \chi_m(x).
\end{eqnarray*}
Here $\prod_{x \in S}$ is the product over all elements in the tuple $S$ (where some elements may occur multiple times). Now observe that the polynomial $P$ corresponding to an update sequence $u_\ell,\dots,u_1$ can be written as the sum of two polynomials $P = Q + R$, where $Q$ is the polynomial corresponding to performing all updates except those in epoch $i$, and $R$ corresponds to performing only the updates of epoch $i$. Since we fixed all updates outside epoch $i$, the polynomial $Q$ is fixed. The polynomial $R$ on the other hand is precisely uniform random over all degree $n_i-1$ polynomials over $\GF(2^d)$. It follows that the evaluations $P(x)$ are $n_i$-wise independent, i.e. for any $h \leq n_i$ distinct elements $x_1,\dots,x_h \in \GF(2^d)$ and any set of (not necessarily distinct) values $y_1,\dots,y_h \in \GF(2^d)$, we have 
$$
\Pr_{\ep_i}\left[ \wedge_{i=1}^h P(x_i) = y_i\right] = 2^{-d h}.
$$
This in particular implies that the entries of $\psi_{\ep_i}$ are uniform random and $n_i$-wise independent. Using this observation, we observe that
$$
\E_{\ep_i} \left[\prod_{x \in S} \psi_{\ep_i}(x) \right]
$$
is $0$ if $k \leq n_i$ and at least one $x$ occurs an odd number of times in $S$. On the other hand, if all $x$ in $S$ occur an even number of times, then both $\E_{\ep_i} \left[\prod_{x \in S} \psi_{\ep_i}(x) \right]$ and $\prod_{x \in S} \chi_m(x)$ are equal to $1$. The number of tuples $S \in \GF(2^d)^k$ with all elements occuring an even number of times is at most $\binom{2^d+k/2-1}{k/2}k!$. We thus have
$$
\E_{\ep_i}[ \langle \psi_{\ep_i}, \chi_m \rangle^k] \leq \binom{2^d+k/2-1}{k/2}k! \leq \left(\frac{2e(2^d+k/2-1)}{k}\right)^{k/2}k^k.
$$
Since $k$ is even, we know that $\langle \psi_{\ep_i}, \chi_m \rangle^k$ is non-negative. Thus we can insert absolute values:
$$
\E_{\ep_i}[ |\langle \psi_{\ep_i}, \chi_m \rangle|^k] \leq \left(\frac{2e(2^d+k/2-1)}{k}\right)^{k/2}k^k.
$$
Using that $\Pr_{\ep_i}[M_{\pi^*}(\ep_i) = m] \geq |\supp(M_{\pi^*}(\ep_i))|^{-1}\eps/2$ by the second proposition of Lemma \ref{lem:findingM}, it further holds that 
$$
\E_{\ep_i}[ |\langle \psi_{\ep_i}, \chi_m \rangle|^k \mid M_{\pi^*}(\ep_i) = m] \leq 2\eps^{-1}\left(\frac{2e(2^d+k/2-1)}{k}\right)^{k/2}k^k |\supp(M_{\pi^*}(\ep_i))|.
$$
By convexity of $x^k$, it follows from Jensen's inequality that
$$
\E_{\ep_i}[ |\langle \psi_{\ep_i}, \chi_m \rangle| \mid M_{\pi^*}(\ep_i) = m]^k \leq 2\eps^{-1}\left(\frac{2e(2^d+k/2-1)}{k}\right)^{k/2}k^k |\supp(M_{\pi^*}(\ep_i))|.
$$
Taking the $k$'th root, we arrive at
$$
\E_{\ep_i}[ |\langle \psi_{\ep_i}, \chi_m \rangle| \mid M_{\pi^*}(\ep_i) = m] \leq \left(2\eps^{-1}\right)^{1/k}\left(\frac{2e(2^d+k/2-1)}{k}\right)^{1/2}k |\supp(M_{\pi^*}(\ep_i))|^{1/k}.
$$
We thus conclude
\begin{eqnarray*}
\eps 2^d \leq \left(2\eps^{-1}\right)^{1/k}\left(\frac{2e(2^d+k/2-1)}{k}\right)^{1/2}k |\supp(M_{\pi^*}(\ep_i))|^{1/k} &\Rightarrow& \\
\eps^2 2^{d} \leq 2\left(2ek(2^d+k/2-1)\right)^{1/2}|\supp(M_{\pi^*}(\ep_i))|^{1/k}.
\end{eqnarray*}
Setting $k = n_i$ (or $n_i-1$ if $n_i$ is odd), we have $k \leq 2^{d/4}$ (since $n_i \leq n_\ell \leq 2^{d/4}$). For this choice of $k$, the above gives us:
\begin{eqnarray*}
\eps^2 2^d \leq 2^{(5/8)d + O(1)} |\supp(M_{\pi^*}(\ep_i))|^{1/n_i}.
\end{eqnarray*}
For $\eps \geq 2^{-d/8}$, this gives
\begin{eqnarray*}
2^{d/8-O(1)} \leq |\supp(M_{\pi^*}(\ep_i))|^{1/n_i} \Rightarrow \\
\log_2\left(|\supp(M_{\pi^*}(\ep_i))|\right) = \Omega(n_i d).
\end{eqnarray*}
Since we assumed the protocol has Alice sending $o(n_i d)$ bits, we conclude that the protocol $\pi$ must have advantage less than $2^{-d/8}$. Since this holds for any protocol, we conclude $\suc\left(G^i_\cP, \;\cU, \; o(n_id)\right) \leq 2^{-d/8}$.

\paragraph{Proof of Lemma~\ref{lem:findingM}.}
For each $m \in \supp(M_{\pi^*}(\ep_i))$, define $z_m$ to take the value $z_m:=1/\Pr_{\ep_i}[M_{\pi^*}(\ep_i) = m]$ and $y_m := 2^d-\E_{\ep_i}[|\langle \psi_{\ep_i} , \chi_{M_{\pi^*}(\ep_i)} \rangle | \mid M_{\pi^*}(\ep_i) = m]$. Both $z_m$ and $y_m$ are non-negative for all $m \in \supp(M_{\pi^*}(\ep_i))$. We observe that $\E_{\ep_i}[z_{M_{\pi^*}(\ep_i)}] = |\supp(M_{\pi^*}(\ep_i))|$. Secondly, we have 
\begin{eqnarray*}
\E_{\ep_i}[y_{M_{\pi^*}(\ep_i)}] &=& 2^d -\E_{\ep'_i}[ \E_{\ep_i}[|\langle \psi_{\ep_i} , \chi_{M_{\pi^*}(\ep_i)} \rangle | \mid M_{\pi^*}(\ep_i) = M_{\pi^*}(\ep'_i) ]] \\
&=& 
2^d -\sum_{m \in \supp(M_{\pi^*}(\ep_i))} \Pr[M_{\pi^*}(\ep_i) = m] \E_{\ep_i}[|\langle \psi_{\ep_i} , \chi_{M_{\pi^*}(\ep_i)} \rangle | \mid M_{\pi^*}(\ep_i) = m]\\
&=& 2^d - \E_{\ep_i}[|\langle \psi_{\ep_i} , \chi_{M_{\pi^*}(\ep_i)} \rangle | ] \\
&\leq& 2^d -  \eps 2^{d+1}.
\end{eqnarray*} 
From Markov's inequality, we conclude that
$$
\Pr[y_{M_{\pi^*}(\ep_i)}  > 2^d - \eps 2^{d}] < \frac{2^d -  \eps 2^{d+1}}{2^d - \eps 2^d} = 1-\frac{\eps 2^d}{2^d - \eps 2^d} \leq 1-2\eps.
$$
Similarly, we conclude
$$
\Pr[z_{M_{\pi^*}(\ep_i)} > \eps^{-1}|\supp(M_{\pi^*}(\ep_i))|/2] < 2\eps.
$$
By a union bound, we conclude that there is some $m \in \supp(M_{\pi^*}(\ep_i))$ satisfying both:
$$
y_m \leq 2^d - \eps 2^{d} \Rightarrow \E_{\ep_i}[|\langle \psi_{\ep_i} , \chi_{M_{\pi^*}(\ep_i)} \rangle | \mid M_{\pi^*}(\ep_i) = m] \geq \eps 2^d.
$$
and
$$
z_{m} \leq \eps^{-1}|\supp(M_{\pi^*}(\ep_i))|/2 \Rightarrow \Pr_{\ep_i}[M_{\pi^*}(\ep_i) = m] \geq |\supp(M_{\pi^*}(\ep_i))|^{-1}\eps/2.
$$

\section{Lower Bound for Parity Searching in Butterfly Graphs}
\label{sec:lowerparity}
In this section, we prove a lower bound for a new and dynamic version of P{\v a}tra{\c s}cu's distance oracles on ``Butterfly" graphs, 
which is the key for the rest of our lower bounds. We start by formally defining the problem, which we call \emph{parity searching in Butterfly graphs}. 


\subsection{Parity Searching in Butterfly Graphs}
\label{sec:paritysearchdef}

\begin{figure}[h]
\centering
\includegraphics[width=6cm, keepaspectratio]{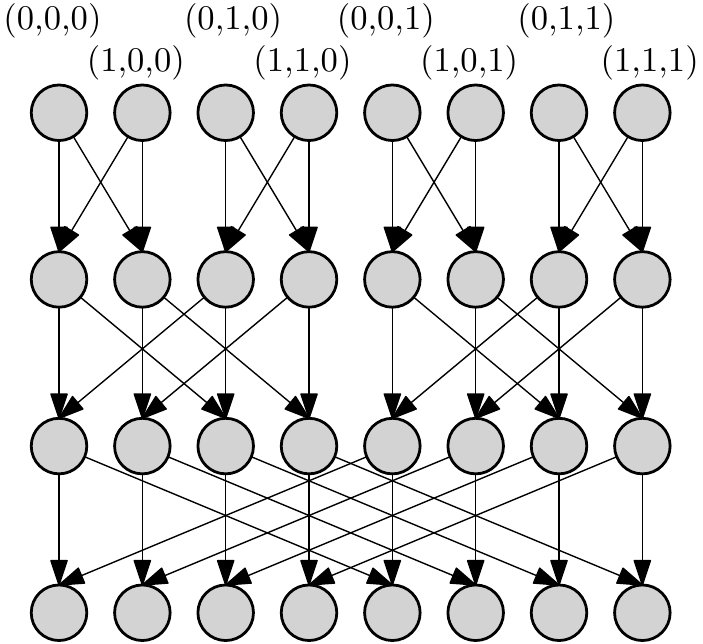}
\caption{The Butterfly graph with degree 2 and depth 3. Nodes are represented as vectors in $[2]^3$, where the $i$'th coordinate corresponds to the $i$'th bit of the binary representation of the column index. Edges leaving a node at level $i$ go to nodes at level $i+1$ whose column index may differ only in the $i$'th bit.}
\label{fig:butterfly}
\end{figure}

To introduce the problem, we first define the Butterfly graph.
The Butterfly graph with degree $B$ and depth $d$ is defined as follows: It has $d+1$ layers, each with $B^d$ vertices. The vertices on level $0$ are sources, while the ones on level $d$ are sinks. Each vertex except the sinks has out-degree $B$, and each vertex except the sources has in-degree $B$. Let $v_0,\dots,v_{B^d-1}$ denote the nodes at some level of the Butterfly. We think of the nodes as being indexed by vectors in $[B]^d$, where the $i$'th coordinate of the vector corresponding to $v_j$ equals the $i$'th bit of $j$. With this representation, there are precisely two edges going out of a node $v_j$ on level $i$, and these two edges go to the two nodes $w_j$ and $w_k$ at level $i+1$ such that $k$ differs from $j$ only on the $i$'th bit (and $j=j$). See Figure~\ref{fig:butterfly} for a Butterfly graph with degree $2$ and depth $3$.

The Butterfly graph has the property that there is a unique source-sink path for each source-sink pair, corresponding to ``morphing'' the binary representation of the source index into the sink index, one bit at a time.

\begin{figure}[h]
\centering
\includegraphics[width=6cm, keepaspectratio]{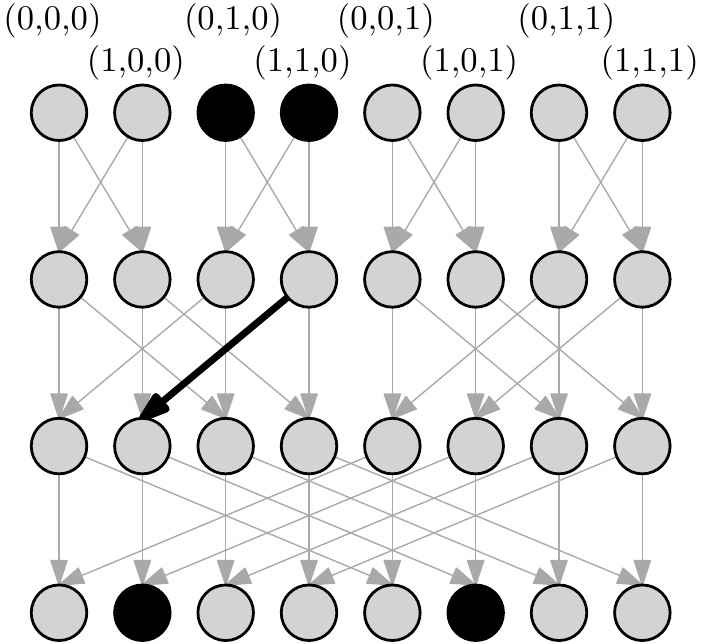}
\caption{The bold edge goes from the level 1 node indexed by $(1,1,0)$ to the level 2 node indexed by $(1,0,0)$. The set of source-sink pairs having their unique path routing through the edge is $(\star,1,0) \times (1,0,\star) = \{(0,1,0),(1,1,0)\} \times \{(1,0,0),(1,0,1)\}$.}
\label{fig:butterflyRoute}
\end{figure}

For a directed edge $(u,v)$ in the Butterfly graph, let $j$ denote
the level of $u$ and $j+1$ the level of $v$. Let $\hat{u} =
(\dots,w_{j-1},w_j, w_{j+1},\dots) \in [B]^d$ denote the
vector corresponding to $u$ and $\hat{v} =  (\dots,w_{j-1},w'_j,
w_{j+1},\dots) \in [B]^d$ denote the vector corresponding to
$v$. The crucial property of the Butterfly graph is that the set of
source-sink pairs $(s,t)$ that have their unique path routing through
the edge $(u,v)$ are precisely those pairs $(s,t)$ where $s \in
(\star,\star,\dots,w_j, w_{j+1},\dots,w_{d-1}) \subseteq [B]^d$
and $t \in (w_0,\dots,w_{j-1},w'_j, \star, \dots) \subseteq
[B]^d$. See Figure~\ref{fig:butterflyRoute}.

P{\v a}tra{\c s}cu studied reachability in the Butterfly graph. More
precisely, one is given as input a subset of the $n = d B^{d+1}$ possible edges of a Butterfly 
graph. One must preprocess the edges into a data structure, such that
given a source-sink pair $(s,t)$, one can output whether $s$ can reach
$t$. 

We modify P{\v a}tra{\c s}cu's reachability problem
such that we can use it in reductions to prove new \emph{dynamic} lower
bounds.

\paragraph{Parity Searching in One Butterfly.}
First consider the following boolean problem on (a single) 
Butterfly graph: Each edge $(u,v)$ of the Butterfly is assigned a
weight $z^{(u,v)}$ amongst $0$ and $1$. A query is defined by a
pair $(s,t) \in [B^d] \times [B^d]$. We think of $s$ as the index of a source node. For $t$, compute the number $\hat{t} = \overleftarrow{t}$ and think of $\hat{t}$ as the index of a sink. The goal is to sum the weights assigned to the set of edges on the unique source-sink path from $s$ to $\hat{t}$.
 Here $\overleftarrow{x}$ of an integer $x\in [B^{d}]$ is the number obtained by writing $x$ in
base $B$ and then reversing the order of digits. That is, if $x =
\sum_{j=0}^{d-1}w_j B^j$ for $w_j \in [B]$, then
$\overleftarrow{x} = \sum_{j=0}^{d-1} w_j B^{d-j-1}$. Note that
$\overleftarrow{x}$ preserves the leading $0$'s and thus when
reversing $t$ we include potential
leading $0$'s in such a way that before reversing, we have precisely
$d$ digits. Reversing the digits of $t$ is an idea by P{\v a}tra{\c s}cu. It has the crucial effect that the set of queries $(s,t)$ summing the weight of an edge $(u,v)$ will correspond to a rectangle. More formally, recall that the source-sink pairs $(s',t') \in [B]^d \times [B]^d$ that route through an edge $(u,v)$ is precisely the set where $s' \in
(\star,\star,\dots,w_j, w_{j+1},\dots,w_{d-1}) \subseteq [B]^d$
and $t' \in (w_0,\dots,w_{j-1},w'_j, \star, \dots) \subseteq
[B]^d$. Translating this back to the query pairs $(s,t) \in [B^d] \times [B^d]$ (where we recall that the sink index is obtained by reversing the digits of $t$), we get that the set of queries that must sum the weight of an edge are those where 
$$
s \in \left[ \sum_{k=j}^{d-1} w_k B^k, \sum_{k=j}^{d-1} w_k B^k + \sum_{k=0}^{j-1} (B-1)B^k \right]
$$
and
$$
t \in \left[ w'_j B^{d-j-1} + \sum_{k=0}^{j-1} w_k B^{d-1-k}, w'_j B^{d-j-1} + \sum_{k=0}^{j-1} w_k B^{d-1-k}  + \sum_{k=0}^{d-j-2} (B-1)B^k\right].
$$
Note that this is a rectangle. This already hints at how we are going to use Butterfly graphs in a reduction to 2D rectangle stabbing.

\begin{figure}[h]
\centering
\includegraphics[width=6cm, keepaspectratio]{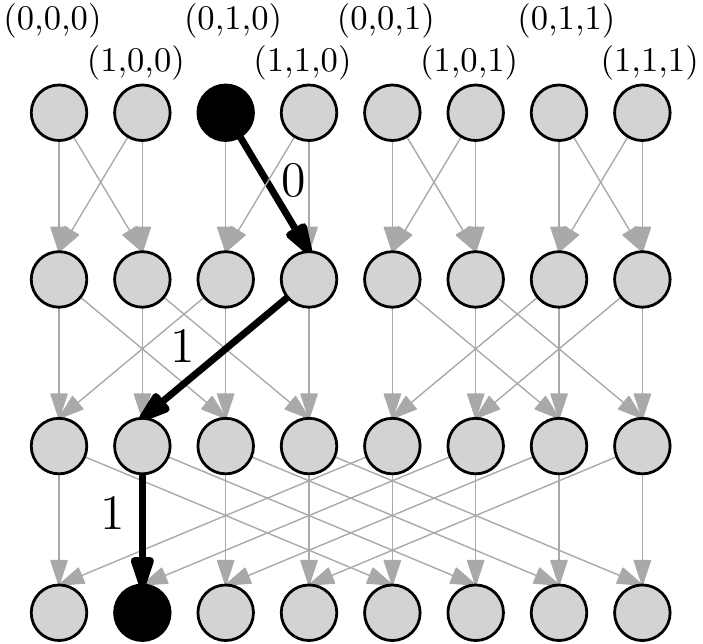}
\caption{Given the query $(2,4)$, we interpret $2$ as the index of a source. The number $4$ is written in binary (with 3 bits) and then reversed, yielding the number $1$. The query $(2,4)$ thus asks to sum the weights of edges on the path from the source indexed by $2 = (0,1,0)$ to the sink indexed by $1 = (1,0,0)$. In this example, two of these edges are assigned the weight $1$, so the XOR of the weights, and thus the answer to the query $(2,4)$, is $0$.}
\label{fig:butterflyQuery}
\end{figure}

Denoting by $p(s,t)$ the unique set of edges
on the path from the source indexed by $s$ to the sink indexed by $\overleftarrow{t}$, the goal is thus to compute $\oplus_{(u,v) \in
  p(s,t)} z^{(u,v)}$, where $\oplus$ denotes XOR (parity). See Figure~\ref{fig:butterflyQuery}.

\paragraph{Parity Searching in Butterfly Graphs.}
We extend
this problem to a dynamic data structure problem, \emph{parity
  searching in Butterfly graphs}, as follows: We have $\ell$ Butterfly
graphs $G_\ell,\dots,G_1$ where all $G_i$'s have the same degree $B$,
but varying depths $d_\ell,\dots,d_1$ such that $d_i = 8i\lg(t_u w)/\lg B$. We will from here and onwards fix $B = (w t_u)^8$ to ensure that depths are integers (they then become $d_i = i$). The idea is that we will have an epoch of updates corresponding to each $G_i$. Since the number of edges in a Butterfly graph of depth $d_i$ is $d_iB^{d_i+1}$, this means that the epoch sizes go down by a factor $\beta = \Theta(B^{8 \lg(t_u w)/\lg B}) = \Theta((t_u w)^8)$. 

Initially all edges of all Butterflies have weight $0$. An update is specified by an index $i \in
\{1,\dots,\ell\}$, an edge $(u,v) \in G_i$ and a weight $y \in
\{0,1\}$. It has the effect of changing the weight of the edge $(u,v)
\in G_i$ to $z^{(u,v)} \gets y$. For technical reasons (for use in reductions), we require that throughout a sequence
of updates, no edge has its weight set more than once. 

A query is specified by two indices $s,t \in [B^{d_\ell}]$ and the answer to the query is:
$$
\bigoplus_{i=1}^\ell \bigoplus_{(u,v) \in p_i(s,t)} z^{(u,v)}
$$
where $p_i(s,t)$ is the set of edges on the path from the source       
$s_i =\lfloor s/B^{d_\ell-d_i} \rfloor$ to the sink $t_i = \overleftarrow{\lfloor  
t/B^{d_\ell-d_i}\rfloor}$ in $G_i$. 

\begin{figure}[h]
\centering
\includegraphics[width=12cm, keepaspectratio]{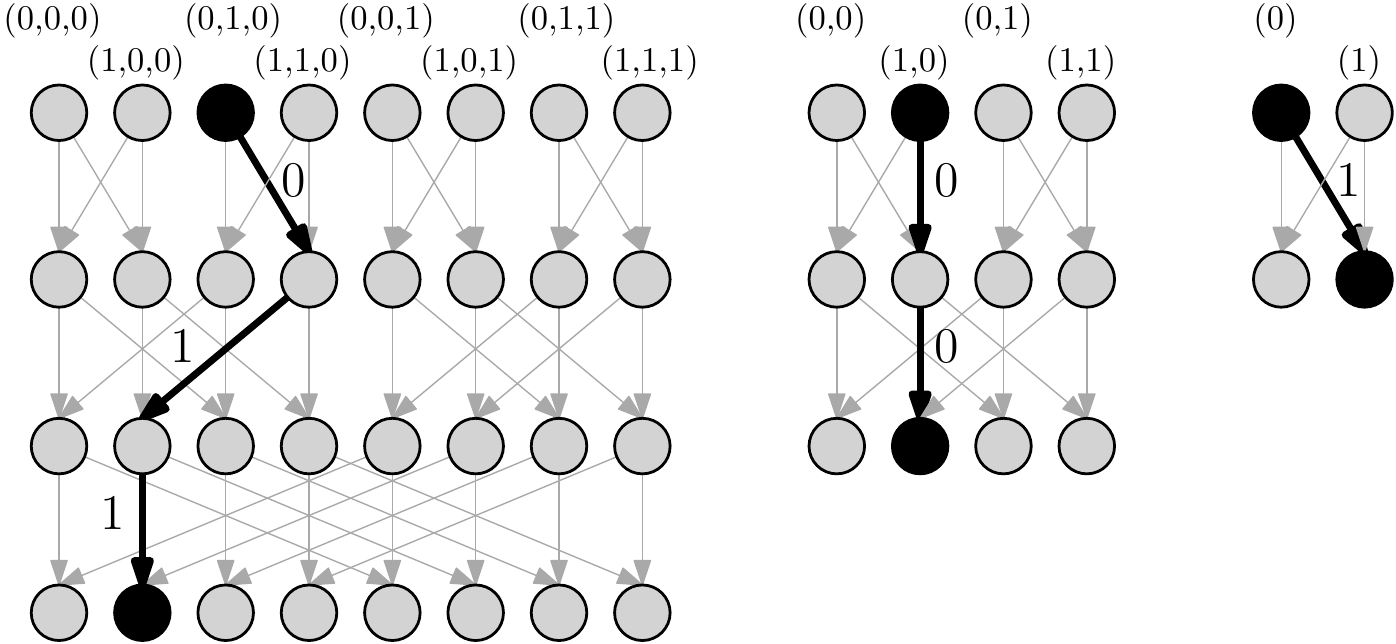}
\caption{Consider parity searching in the three Butterfly graphs of degree $2$ and depths $3,2$ and $1$ respectively. Given the query $(2,4)$, we interpret $\lfloor 2/2^0 \rfloor = 2$ as the index of a source in the first graph, $\lfloor 2/2^1 \rfloor = 1$ as the source index in the second and $\lfloor 2/2^2 \rfloor = 0$ as the index of the source in the third and smallest graph. The number $\lfloor 4 /2^0 \rfloor = 4$ is written in binary (with 3 bits) and then reversed, yielding the sink indexed $1$ in the first graph. Similarly $\lfloor 2 / 2^1 \rfloor = 2$ is written in binary with 2 bits and reversed, yielding the sink indexed $1$ in the second graph. Finally, $\lfloor 4/2^2 \rfloor = 1$ is written in binary with 1 bit and reversed, yielding the sink indexed $1$ in the third and smallest graph. The query $(2,4)$ thus asks to sum the weights of edges along all the paths in bold, yielding the result $0+1+1+0+0+1 = 1$ (we are computing parity).}
\label{fig:butterflyMany}
\end{figure}

To summarize a query in words, the indices $s$ and $t$ are
translated to a set of source-sink pairs, one in each $G_i$, by integer division
with $B^{d_\ell-d_i}$ and then reversing the bits of the number
obtained from $t$. We must then compute the parity of the weights
assigned to all the edges along the $\ell$ corresponding paths
$p_1(s,t),\dots,p_\ell(s,t)$. Note that the total input size is $ n =
\sum_{i=1}^\ell d_i B^{d_i+1} = O(d_\ell B^{d_\ell + 1})$ and that the
size of consecutive graphs $G_i$ and $G_{i-1}$ differ by a factor
$(t_u w)^{\Theta(1)}$. See Figure~\ref{fig:butterflyMany} for an example.

\subsection{The Lower Bound}
\label{sec:paritysearchdef}

Let $\cP$ denote the dynamic problem of parity searching in Butterfly graphs. Recall that in $\cP$, we have 
$\ell$ Butterfly graphs $G_\ell,\dots,G_1$ where all $G_i$'s have the same degree $B = (w t_u)^8$, but varying depths $d_\ell,\dots,d_1$ such that $d_i = 8i\lg(t_u w)/\lg B = i$.

\paragraph{Hard Distribution.}
We consider the following random sequence of updates $\cU = (\ep_\ell,\ep_{\ell-1},\ldots,\ep_1)$, where $|\ep_i|=n_i = d_iB^{d_i+1}$. The updates $\ep_i$ of epoch $i$ assign a uniform random weight amongst $\{0,1\}$ to each edge of the Butterfly $G_i$ (which has precisely $d_i B^{d_i+1}$ edges). The query $q = (s,t)$ has $s$ and $t$ drawn independently and uniformly at random from $[B^{d_\ell}]$.

\subsubsection{Meta Queries}
Let $D$ be a dynamic data structure for parity searching in Butterfly graphs of degree $B$, having worst case update time $t_u$  and expected query time $t_q$ under $\cU$. We 
cannot apply Theorem~\ref{thm_weak_simulation_informal} directly to this problem by proving a strong lower bound for the possible advantage on epochs $i \in [\ell/2,\ell]$. 
In fact, there is a randomized one-way protocol achieving advantage $1/\poly(\lg n)$ with $n_i/\poly(\lg n)$ communication for any epoch $i$. To get the lower bound we are after, we need to show that with $n_i/\poly(\lg n)$ communication, the best achievable advantage is only $1/\poly(n)$. To ensure the latter, we need to perform certain technical 
manipulations on queries of $\cP$ in our simulation. We do this as follows:

First, by arguments similar to those in the proof of Theorem~\ref{thm_weak_simulation_informal}, there must be an epoch $i \in [\ell/2, \ell]$ such that
$$
\frac{1}{|\cQ|} \sum_{q \in \cQ} \E[T^i_q] \leq 2t_q/\ell.
$$
Recall that $T^i_q$ is the number of cells associated to epoch $i$ which is probed by $D$ when answering $q$ (see Section~\ref{sec:simulation}). Fix such an epoch $i$.

Recall that a query in $\cQ$ is specified by a tuple $(s,t)$ with $s,t \in [B^{d_\ell}]$. Each such tuple corresponds to a source-sink path by integer division with $B^{d_\ell-d_i}$ and reversing the digits of the number obtained from $t$. Hence there are exactly $B^{d_\ell-d_i}$ values of $s$ that specify the same source. Likewise for the sinks.

\begin{figure}[h]
\centering
\includegraphics[width=6cm, keepaspectratio]{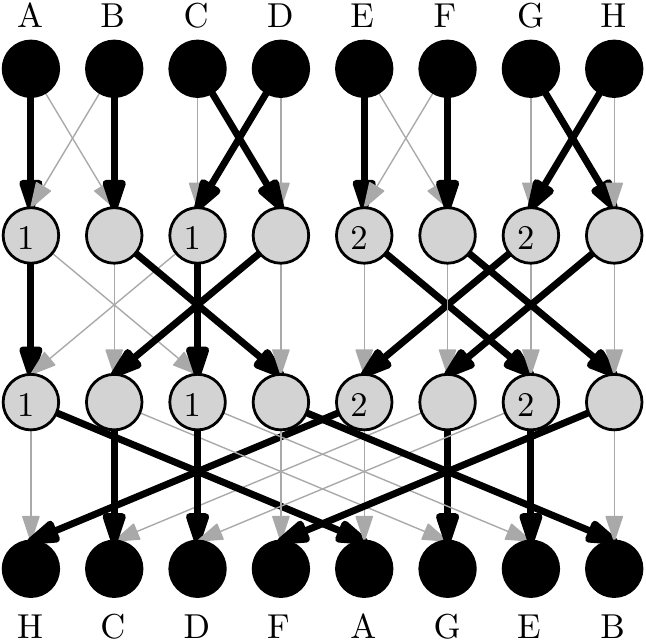}
\caption{Two chunks have been marked in this figure, labeled ``1'' and ``2'' respectively. In each chunk, a permutation on two elements define two edges, one leaving each node at level $1$ and one entering each node at level $2$. The permutation assigned to chunk ``1'' is the permutation $\pi(0)=0, \pi(1)=1$, whereas the permutation assigned to chunk ``2'' is $\pi(0)=1, \pi(1)=0$. Such a permutation is assigned to every chunk of the graph. We can then trace a path from each source down to a sink, resulting in $8$ node-disjoint source-sink paths. The edges on these paths are shown in bold and the corresponding source-sink pairs are labeled such that the source labeled ``A'' is paired with the sink labeled ``A'' and so forth.}
\label{fig:butterflyMeta}
\end{figure}

From the set of source-sink paths, we define a collection of meta queries $\cQ^*$. For each level $j$ of the Butterfly graph $G_i$, recall that edges go between vertices whose base-$B$ vector differ only in the $j$'th coordinate. We thus group vertices in level $j$ and $j+1$ into $B^{d_i-1}$ chunks. Each chunk consists of all vertices in level $j$ and $j+1$ whose corresponding vectors agree in all but the $j$'th coordinate, i.e. a chunk consists of all vertices with the corresponding vector being $(w_0,\dots,w_{j-1}, \star, w_{j+1},\dots,w_{d_i-1})$. Summed over all $d_i$ levels with outgoing edges, we have $d_iB^{d_i-1}$ chunks. Now consider assigning a permutation on $B$ elements to each chunk. Using $\pi_j^{(w_0,\dots,w_{j-1}, \star, w_{j+1},\dots,w_{d_i-1})}$ to denote the permutation at level $j$ corresponding to vertices in level $j$ and $j+1$ with vectors of the form $(w_0,\dots,w_{j-1}, \star, w_{j+1},\dots,w_{d_i-1})$, such an assignment of permutations to all chunks now yield $B^{d_i}$ unique source-sink pairs as follows: For each source $s' = (w_0,\dots,w_{d_i-1})$, trace a path as follows: Start by going to the level $1$ vertex with vector $(w_0'=\pi_0^{(\star,w_1\dots,w_{d_i-1})}(w_0), w_1,\dots, w_{d_i-1})$. Then to the level $2$ vertex with vector $(w_0',w_1' =\pi_0^{(w'_0,\star,w_2\dots,w_{d_i-1})}(w_1),\dots, w_{d_i-1})$ and so forth until we reach a sink $t' = (w_0',\dots,w_{d_i-1}')$. Since we use a permutation in each chunk, the set of $B^{d_i}$ constructed source-sink pairs have the property that \emph{exactly} one path passes through each vertex at each level. See Figure~\ref{fig:butterflyMeta}.

Now consider such a collection of $B^{d_i}$ source-sink pairs $(0,t_1),\dots,(B^{d_i}-1,t_{B^{d_i}-1})$ (recall there is precisely one query per source and one per sink). We create a number of
meta queries for each such collection of $B^{d_i}$ source-sink pairs. The set of meta queries corresponding to source-sink pairs $(0,t_0),\dots,(B^{d_i}-1,t_{B^{d_i}-1})$ is precisely the set of all query lists $S=(s'_0,t'_0),\dots,(s'_{B^{d_i}-1}, t'_{B^{d_i}-1})$ with $\lfloor s'_j/B^{d_\ell-d_i} \rfloor = j$ and $\overleftarrow{\lfloor t'_j/B^{d_\ell-d_i} \rfloor} = t_j$ for all $j=0,\dots,B^{d_i}-1$. The meta query $q^S$ then has the answer 
$$
\cA(q^S) := \bigoplus_{(s'_j,t'_j) \in S} \bigoplus_{i=1}^\ell \bigoplus_{(u,v) \in p_i(s'_j,t'_j)} z^{(u,v)}
$$
Said in words, the meta query $q^S$ asks to compute the XOR of the answer to all the queries in $S$. Note that all paths involved in the XORs are disjoint in $G_i$, and thus the weight of any edge in $G_i$ is included at most once in this sum. We use $\cQ^*$ to denote the collection of all meta queries.

Now the idea is that we can use $D$ to answer such meta queries efficiently. More specifically, consider running the same distribution $\cU$ of updates, but instead of drawing the query tuple $q$ as above, we instead draw a uniform random meta query $q^S$ and ask to output its answer. Call the resulting dynamic data structure problem $\cP^*_i$. We can use the data structure $D$ to obtain an efficient solution for this problem: When receiving the meta query $q^S$, we simply run the query algorithm for each $(s_j,t_j) \in S$ and compute the resulting XOR of query answers. Clearly this gives the correct result. Now the crucial observation is that if we draw a uniform random query $(s_j,t_j)$ from $q^S$, then the distribution of that query is still uniform over all queries to the \emph{original} problem, i.e. the distribution of $(s_j,t_j)$ is simply a uniform random tuple in $[B^{d_\ell}] \times [B^{d_\ell}]$. Thus by linearity of expectation, we have:
$$
\frac{1}{|\cQ^*|} \sum_{q^S \in \cQ^*} \E[T^i_{q^S}]= B^{d_i}\frac{1}{|\cQ|} \sum_{q \in \cQ}  \E[T^i_{q}] \leq 2B^{d_i}t_q/\ell.
$$
Here $T^i_{q^S}$ is the number of cells associated to epoch $i$ that is probed when answering $q^S$ in the above manner. We also have:
$$
\frac{1}{|\cQ^*|} \sum_{q^S \in \cQ^*} \E[T^i_{q^S}] = B^{d_i}t_q.
$$
We now wish to invoke Theorem~\ref{thm_weak_simulation_epoch}. The theorem requires $t_q (w t_u)^{a+1} \leq n_i$ for a constant $a>1$. The average expected query time for a meta query was $B^{d_i}t_q$, thus we must have $B^{d_i} t_q (w t_u)^{a+1} \leq n_i$ for a constant $a>1$. Since $n_i = d_i B^{d_i+1}$, we see that it suffices to have $B \geq t_q (w t_u)^{a+1}$. We can assume $t_q \leq \lg^2n$, as otherwise we have already finished our proof. Therefore, we see that any $B = \Omega((w t_u)^{a+3})$ suffices (as $w = \Omega(\lg n)$). We chose $B = (w t_u)^8$, so we can apply the theorem with any $a \leq 5$. Furthermore, the epoch sizes go down by a factor $\beta = \Theta((w t_u)^8)$ which also satisfies the requirements of the theorem for any choice of $a \leq 8$.

We can thus invoke Theorem~\ref{thm_weak_simulation_epoch}, with $a=2$, to conclude that
\[ \suc\left(G^i_{\cP^*_i}, \;\cU, \; n_i/(w t_u) \right) \geq  \exp\left(-t_qB^{d_i} \log^2 (w t_u)/\sqrt{\log n}\right) . \]
The next section therefore aims to bound the best achievable advantage for this $G^i_{\cP^*_i}$ problem. 

Before we proceed, a few remarks are in order. As discussed earlier, the base problem of answering just one source-sink pair admits a too efficient communication protocol. 
As we shall see in the following subsection, forcing the data structure to answer \emph{multiple} structured queries on the same input (i.e., meta queries) rules out such efficient protocols. Unfortunately, the number of queries we need in a meta query depends on the epoch size. Therefore, we had to \emph{first} zoom in on an epoch $i$ for which the data structure is efficient, and \emph{then} define the meta queries after having chosen the epoch. This is also the reason why we needed the more specific Theorem~\ref{thm_weak_simulation_epoch} rather than Theorem~\ref{thm_weak_simulation_informal}.

The next subsection proves the following result:

\begin{lemma}
\label{lem:noAdvParity}
For any epoch $i \in [\ell/2,\ell]$, we have $\suc\left(G^i_{\cP^*_i}, \;\cU, \; o(n_i)\right) \leq  2^{-(d_i B^{d_i})/8}$.
\end{lemma}

Let us first see why this lemma implies the desired lower bound. We see that it must be the case that
\begin{eqnarray*}
\exp\left(-t_qB^{d_i} \log^2 (w\cdot t_u)/\sqrt{\log n}\right) &\leq& 2^{-(d_i B^{d_i})/8} \Rightarrow \\
t_qB^{d_i} \log^2 (w\cdot t_u)/\sqrt{\log n} &=& \Omega\left(d_i B^{d_i}\right) \Rightarrow \\
t_q &=& \Omega(d_i \sqrt{\log n} / \log^2 (w\cdot t_u))
\end{eqnarray*}
But $d_i \geq d_\ell/2$ for any $i \geq \ell/2$ and $d_\ell = \Omega(\lg_B n)$, thus we conclude that
$$
t_q = \Omega\left(\frac{\log^{3/2} n}{\lg B \log^2 (w\cdot t_u)} \right).
$$
We have set $B = (w t_u)^8$, thus the lower bound becomes:
$$
t_q = \Omega\left(\frac{\log^{3/2} n}{\log^3 (w\cdot t_u)} \right).
$$
This completes the proof of Theorem~\ref{thm:paritysearch}. The next section proves the necessary Lemma~\ref{lem:noAdvParity}.

\subsubsection{Low Advantage on Epochs}
Let $i \in [\ell/2,\ell]$. In the communication game $G^i_{\cP^*_i}$ for the random update sequence $\cU$, Alice receives all updates of all epochs and Bob receives all updates of all epochs except $i$. Bob also receives a meta query $q^S$. Let $\pi$ be a one-way randomized protocol in which Alice sends $o(n_i)$ bits to Bob, and suppose that $\pi$ achieves an advantage of $\eps$ w.r.t. $\ep$ and $q^S$. Since the query $q^S$ and the updates $\ep_i$ are independent of the updates of epochs $\ep_\ell,\cdots,\ep_{i+1},\ep_{i-1},\dots,\ep_1$, we can fix the random coins of the protocol and fix the updates of all epochs except epoch $i$, such that for the resulting deterministic protocol $\pi^*$ and fixed update sequences $u_\ell,\dots,u_{i+1},u_{i-1},\dots,u_1$ we have that Alice never sends more than $o(n_i)$ bits and $\Pr_{\ep_i,q^S}[v_{\pi^*} = \cA(q^S)] \geq 1/2+\eps$. Here $\cA(q^S)$ is the answer to query $q^S$ on updates $u_\ell,\dots,u_{i+1},\ep_i,u_{i-1},\dots,u_1$ and query $q^S$. The random variable $v_{\pi^*}$ is Bob's output when running the deterministic protocol $\pi^*$ on $u_\ell,\dots,u_{i+1},\ep_i,u_{i-1},\dots,u_1$ and query $q^S$.

Let $M_{\pi^*}(\ep_i)$ be the message sent by Alice in procotol $\pi^*$ on updates $u_\ell,\dots,u_{i+1},\ep_i,u_{i-1},\dots,u_1$. Then $v_{\pi^*} = v_{\pi^*}(M_{\pi^*}(\ep_i),q^S)$ is determined from $M_{\pi^*}(\ep_i)$ and $q^S$ alone (since the updates of other epochs are fixed). For each of the possible messages $m$ of Alice, define the vector $\chi_m$ having one coordinate per $q^S \in \cQ^*$. The coordinate $\chi_m(x)$ corresponding to some $q^S$ has the value $-1$ if $v_{\pi^*}(m,q^S)=0$ and it has the value $1$ otherwise. Similarly define for each sequence of updates $u_i \in \supp(\ep_i)$ the vector $\psi_{u_i}$ having one coordinate $\psi_{u_i}(q^S)$ per $q^S \in \cQ^*$, where the coordinate corresponding to some $x$ takes the value $-1$ if the correct answer to the query $q^S$ is $0$ after the update sequence $u_\ell,\dots,u_{i+1},u_i,u_{i-1},\dots,u_1$ and taking the value $1$ otherwise. Since $\pi^*$ has advantage $\eps$ and $q^S$ is uniform in $\cQ^*$, we must have
$$
\E_{\ep_i}[\langle \psi_{\ep_i} , \chi_{M_{\pi^*}(\ep_i)} \rangle ] = ((1/2+\eps) - (1/2-\eps))|\cQ^*| = 2 \eps |\cQ^*|.
$$
This in particular implies that if we take the absolute value of the inner product, we have
$$
\E_{\ep_i}[|\langle \psi_{\ep_i} , \chi_{M_{\pi^*}(\ep_i)} \rangle | ] \geq 2 \eps |\cQ^*|.
$$
By arguments identical to those in the proof of Lemma~\ref{lem:findingM}, we conclude that there must be some $m \in \supp(M_{\pi^*}(\ep_i))$ such that we have both
\begin{itemize}
\item $\E_{\ep_i}[|\langle \psi_{\ep_i} , \chi_{M_{\pi^*}(\ep_i)} \rangle | \mid M_{\pi^*}(\ep_i) = m] \geq \eps|\cQ^*|.$
\item $\Pr_{\ep_i}[M_{\pi^*}(\ep_i) = m] \geq |\supp(M_{\pi^*}(\ep_i))|^{-1}\eps/2.$
\end{itemize}
Consider such an $m$ and the corresponding vector $\chi_m$. We examine the following expectation for an even integer $k$ to be determined:
\begin{eqnarray*}
\E_{\ep_i}[ \langle \psi_{\ep_i}, \chi_m \rangle^k] &=& \\
\sum_{T \in (\cQ^*)^k} \E_{\ep_i} \left[\prod_{q^S \in T} \psi_{\ep_i}(q^S) \chi_m(q^S)\right] &=& \\
\sum_{T \in (\cQ^*)^k} \E_{\ep_i} \left[\prod_{q^S \in T} \psi_{\ep_i}(q^S) \right] \prod_{q^S \in T} \chi_m(q^S).
\end{eqnarray*}
Now recall that $\ep_i$ assigns a uniform random and independently chosen weight in $\{0,1\}$ to each edge of $G_i$. For an edge $(u,v)$, let $y^{(u,v)}$ take the value $1$ if $z^{(u,v)}=0$ and let it take the value $-1$ otherwise. Then 
$$
\psi_{\ep_i}(q^S)  = \prod_{(s'_j,t'_j) \in S} \prod_{i=1}^\ell \prod_{(u,v) \in p_i(s'_j,t'_j)} y^{(u,v)}
$$
It follows that if there is even a single edge $(u,v)$ in $G_i$, such that $(u,v)$ occurs an odd number of times when summed over all $p_i(s'_j,t'_j)$ in all $q^S$ in $T$, then $\E_{\ep_i} \left[\prod_{q^S \in T} \psi_{\ep_i}(q^S) \right] = 0$. If all edges in $G_i$ occur an even number of times, then  $\E_{\ep_i} \left[\prod_{q^S \in T} \psi_{\ep_i}(q^S) \right] \in \{-1,+1\}$, depending on the (fixed) weights assigned to edges in epochs different from $i$. Denoting by $\Gamma$ the number of sets $T \in  (\cQ^*)^k$ such that all edges in 
$G_i$ occur an even number of times in the corresponding source-sink paths, we conclude that 
$$
\E_{\ep_i}[ \langle \psi_{\ep_i}, \chi_m \rangle^k] \leq \Gamma.
$$
Since we assume $k$ is even, we may insert absolute values:
$$
\E_{\ep_i}[ |\langle \psi_{\ep_i}, \chi_m \rangle|^k] \leq \Gamma.
$$
To bound $\Gamma$, consider drawing $k$ meta queries $q^{S_1},q^{S_2},\dots,q^{S_k}$ independently and uniformly at random. We wish to bound the probability that every edge in $G_i$ is included an even number of times when summed over the $k$ meta queries $q^{S_1},\dots,q^{S_k}$. For this, zoom in on a chunk of the butterfly graph between some levels $j$ and $j+1$. Each $q^{S_h}$ assigns a uniform random permutation $\pi_h$ on $B$ elements to the chunk. That each edge occurs an even number of times in the chunk is equivalent to every pair of indices $x,y \in [B]$ satisfying that there is an even number of permutations amongst $\pi_1,\dots,\pi_k$ for which $\pi_h(x)=y$. Call this event $E$. To bound the probability of $E$, first observe that $\pi_1,\dots,\pi_k$ is uniform random amongst $(B!)^k$ possible lists of permutations. We wish to bound the number of such lists $\rho_1,\dots,\rho_k$ that satisfy 
\begin{enumerate}
\item For every pair $(x,y) \in [B] \times [B]$, there is an even number of $\rho_h$ for which $\rho_h(x)=y$.
\end{enumerate}
We upper bound the number of such lists $\rho_1,\dots,\rho_k$ via an encoding argument. Let $\rho_1,\dots,\rho_k$ satisfy 1. We can encode $\rho_1,\dots,\rho_k$ as follows: First specify for every pair $x,y$ how many $\rho_h$ that satisfy $\rho_h(x)=y$. Letting $\Delta(x,y)$ denote this number for a pair $(x,y)$, we encode the $\Delta(x,y)$'s efficiently as follows: First divide each $\Delta(x,y)$ by $2$. The resulting values are still integer since each $\Delta(x,y)$ is even by assumption. Next observe that $\sum_{(x,y)} \Delta(x,y)/2 = kB/2$ as each $\rho_h$ adds $B$ to $\sum_{(x,y)} \Delta(x,y)$. Thus we need to specify a sequence of $B^2$ non-negative integers that sum to $kB/2$. It is well known that the number of such integer sequences is $\binom{B^2 + kB/2-1}{kB/2}$, thus we can specify all $\Delta(x,y)$'s using a total of $\lg \binom{B^2 + kB/2-1}{kB/2}$ bits. Finally, for each pair $(x,y)$ in lexicographic order (i.e. first ordered by $x$, then by $y$), if $\Delta(x,y) > 0$, append $\Delta(x,y)\lg k$ bits to the encoding, specifying the set of $\Delta(x,y)$ indices $h$ amongst $\{1,\dots,k\}$ that have $\rho_h(x)=y$. Clearly the set $\rho_1,\dots,\rho_k$ can be recovered from this encoding. The number of bits used is upper bounded by $\lg \binom{B^2 + kB/2-1}{kB/2} + kB \lg k$, meaning that the number of distinct $\rho_1,\dots,\rho_k$ is upper bounded by $\binom{B^2 + kB/2-1}{kB/2} k^{kB}$. We therefore conclude that
\begin{eqnarray*}
\Pr[E] &\leq& \frac{\binom{B^2 + kB/2-1}{kB/2} k^{kB}}{(B!)^k} \\
&\leq&  \frac{ \left( \frac{e(B^2 + kB/2)}{kB/2} \right)^{kB/2} k^{kB}}{B^{kB}} \\
&=& \frac{ \left( 2e(B^2 + kB/2)\right)^{kB/2} k^{kB}}{B^{kB} (kB)^{kB/2}} 
\end{eqnarray*} 
We now fix $k = B/16$ (we assume $k$ is a even and remark that this can always be achieved by blowing up $w$ and $t_u$ by constant factors to ensure that $B = (w t_u)^8$ is a power of $2$). For this value of $k$, the above is bounded by:
\begin{eqnarray*}
\Pr[E] &\leq& \frac{ \left( 8B^2\right)^{kB/2} k^{kB}}{B^{kB} (16 k^2)^{kB/2}} \\
&=& 2^{-kB/2} = 2^{-B^2/32}.
\end{eqnarray*}

Next recall that the permutations assigned to the chunks are independent. Using the fact that there are $d_i B^{d_i-1}$ chunks, we conclude that the probability that all edges occur an even number of times in $q^{S_1},q^{S_2},\dots,q^{S_k}$ is less than $2^{-(B^2/32)d_i B^{d_i-1}}$. Therefore, we get that
$$
\Gamma \leq |\cQ^*|^k (1/2)^{(d_i B^{d_i+1})/32}.
$$
Using that $\Pr_{\ep_i}[M_{\pi^*}(\ep_i) = m] \geq |\supp(M_{\pi^*}(\ep_i))|^{-1}\eps/2$, we conclude that
$$
\E_{\ep_i}[ |\langle \psi_{\ep_i}, \chi_m \rangle|^k \mid M_{\pi^*}(\ep_i) = m] \leq 2\eps^{-1}|\cQ^*|^k (1/2)^{(d_i B^{d_i+1})/32}|\supp(M_{\pi^*}(\ep_i))|.
$$
By convexity of $x^k$ and Jensen's inequality, we have
$$
\E_{\ep_i}[ |\langle \psi_{\ep_i}, \chi_m \rangle| \mid M_{\pi^*}(\ep_i) = m]^k \leq 2\eps^{-1}|\cQ^*|^k (1/2)^{(d_i B^{d_i+1})/32}|\supp(M_{\pi^*}(\ep_i))|.
$$
Taking the $k$'th root, we get
$$
\E_{\ep_i}[ |\langle \psi_{\ep_i}, \chi_m \rangle| \mid M_{\pi^*}(\ep_i) = m] \leq (2\eps^{-1})^{1/k}|\cQ^*| (1/2)^{(d_i B^{d_i+1})/(32k)}|\supp(M_{\pi^*}(\ep_i))|^{1/k}.
$$
Using that $\E_{\ep_i}[|\langle \psi_{\ep_i} , \chi_{M_{\pi^*}(\ep_i)} \rangle | \mid M_{\pi^*}(\ep_i) = m] \geq \eps|\cQ^*|$, we conclude that
\begin{eqnarray*}
\eps|\cQ^*| &\leq& (2\eps^{-1})^{1/k}|\cQ^*| (1/2)^{(d_i B^{d_i+1})/(32k)}|\supp(M_{\pi^*}(\ep_i))|^{1/k} \Rightarrow \\
\eps^{2} &\leq& 2 (1/2)^{(d_i B^{d_i+1})/(32k)}|\supp(M_{\pi^*}(\ep_i))|^{1/k}.
\end{eqnarray*}
Assuming $\eps \geq (1/2)^{(d_i B^{d_i+1})/(128k)}$, we must have:
\begin{eqnarray*}
|\supp(M_{\pi^*}(\ep_i))|^{1/k} &\geq& 2^{(d_i B^{d_i+1})/(64k)-1}.
\end{eqnarray*}
Taking logs, we conclude that
\begin{eqnarray*}
\log_2\left(|\supp(M_{\pi^*}(\ep_i))|\right) = \Omega(d_i B^{d_i+1}) = \Omega(n_i).
\end{eqnarray*}
The contrapositive says that assuming $\log_2\left(|\supp(M_{\pi^*}(\ep_i))|\right) = o(n_i)$, we must have $\eps \leq 2^{-(d_i B^{d_i+1})/(128k)}$. Inserting $k = B/16$, this becomes $\eps \leq 2^{-(d_i B^{d_i})/8}$.

\section{Reductions}
\label{sec:reductions}
This section presents the reductions used to prove the various lower bounds discussed in Section~\ref{sec:concreteproblems}.

\subsection{2D Range Parity and 2D Rectangle Stabbing}
\label{sec:rangeparity}
As argued in Section~\ref{sec:concreteproblems}, a folklore reduction shows that 2D range counting and 2D rectangle stabbing are equivalent problems. The same goes for 2D range parity and 2D rectangle parity, where 2D rectangle parity is the version of rectangle stabbing where we need only return the parity of the number of rectangles containing the query point $q$. Thus we start by giving a reduction from parity searching in Butterfly graphs to 2D rectangle parity.

Recall that in \emph{parity searching in Butterfly graphs} (see Section~\ref{sec:paritysearchdef}), we have $\ell$ Butterfly graphs $G_\ell,\dots,G_1$ where all $G_i$'s have the same degree $B$, but varying depths $d_\ell,\dots,d_1$. Initially all edges of all Butterflies have weight $0$. An update is specified by an index $i \in \{1,\dots,\ell\}$, an edge $(u,v) \in G_i$ and a weight $y \in \{0,1\}$. It has the effect of changing the weight of the edge $(u,v) \in G_i$ to $z^{(u,v)} \gets y$. A query is specified by two indices $s,t \in [B^{d_\ell}]$ and the answer to the query is:
$$
\bigoplus_{i=1}^\ell \bigoplus_{(u,v) \in p_i(s,t)} z^{(u,v)}
$$
where $p_i(s,t)$ is the set of edges on the path from the source $s_i =
\lfloor s/B^{d_\ell-d_i} \rfloor$ to the sink $t_i = \overleftarrow{\lfloor
t/B^{d_\ell-d_i}\rfloor}$ in $G_i$.

We want to show that we can solve this problem using a dynamic data structure for 2D rectangle parity. To do this, we associate a rectangle with each edge $(u,v)$ in some $G_i$. Let $j$ denote
the level of $u$ and $j+1$ the level of $v$. Recall the notation from Section~\ref{sec:concreteproblems} and let $\hat{u} =
(\dots,w_{j-1},w_j, w_{j+1},\dots) \in [B]^{d_i}$ denote the
vector corresponding to $u$ and $\hat{v} =  (\dots,w_{j-1},w'_j,
w_{j+1},\dots) \in [B]^{d_i}$ denote the vector corresponding to
$v$. The crucial property of the Butterfly graph is that the set of
source-sink pairs $(s_i,t_i)$ in $G_i$ that have their unique path routing through
the edge $(u,v)$ are precisely those pairs $(s_i,t_i)$ where $s_i \in
(\star,\dots,\star, w_j, w_{j+1},\dots,w_{d_i-1}) \subseteq [B]^{d_i}$
and $t_i \in (w_0,\dots,w_{j-1},w'_j, \star, \dots) \subseteq
[B]^{d_i}$. 

Now recall that from a query $(s,t) \in [B^{d_\ell}] \times [B^{d_\ell}]$ to parity searching in Butterfly graphs, we must sum the weights along the paths between the source-sink pairs $s_i = \lfloor s/B^{d_\ell-d_i} \rfloor$ and $t_i = \overleftarrow{\lfloor t/B^{d_\ell-d_i} \rfloor}$ in $G_i$ for each $i=1,\dots,\ell$. If we write $(s,t) \in [B^{d_\ell}] \times [B^{d_\ell}]$ as the vectors $s = (w^s_0,\dots,w^s_{d_\ell-1})$ and $t = (w^t_0,\dots,w^t_{d_\ell-1})$ where $w_0^s$ and $w^t_0$ are the least significant digits of $s$ and $t$ in base $B$, then we have that $s_i = (w^s_{d_\ell-d_i},\dots,w^s_{d_\ell-1})$ and $t_i = (w^t_{d_\ell-1},\dots,w^t_{d_\ell-d_i})$. We conclude that the weight of the edge $(u,v)$ in $G_i$ must be counted iff 
\begin{itemize}
\item $w^s_{k+(d_\ell - d_i)} = w_k$ for all $k=j,\dots,d_i-1$.
\item $w^t_{d_\ell-1-k} = w_k$ for $k=0,\dots,j-1$.
\item $w^t_{d_\ell-j-1} = w'_j$.
\end{itemize}
But $s$ and $t$ are integers in $[B^{d_\ell}]$ and the above requirements are thus captured precisely by:
\begin{itemize}
\item $s \in \left[ \sum_{k=j}^{d_i-1} w_k B^{k+(d_\ell - d_i)} , \sum_{k=j}^{d_i-1} w_k B^{k+(d_\ell - d_i)} + \sum_{k=0}^{j-1 + (d_\ell-d_i)}(B-1)B^k \right]$.
\item $t \in \left[w'_j B^{d_\ell-j-1} + \sum_{k=0}^{j-1} w_k B^{d_\ell-1-k} , w'_j B^{d_\ell-j-1} + \sum_{k=0}^{j-1} w_k B^{d_\ell-1-k}  +  \sum_{k=0}^{d_\ell-j-2}(B-1)B^k\right]$.
\end{itemize}
This is equivalent to the point $(s,t)$ being inside a 2D rectangle that depends only on $(u,v)$ and $i$. We denote this rectangle by $R^{(u,v)}_i$. Our reduction now goes as follows: On an update setting the weight of an edge $(u,v)$ in $G_i$ to $0$, we do nothing. If the update sets the weight to $1$, we insert the rectangle $R^{(u,v)}_i$. To answer a query $(s,t)$ to parity searching in Butterfly graphs, we simply ask the 2D rectangle parity query $(s,t)$. Correctness follows immediately by the above arguments and the fact that the definition of parity searching in Butterfly graphs from Section~\ref{sec:concreteproblems} requires that any edge has its weight set at most once. To the worried reader, note that translating an edge $(u,v)$ in $G_i$ to $R^{(u,v)}_i$ does not require any memory lookups, it is purely computational. Thus in the cell probe model, the translation is free of charge.

For completeness, we also sketch the reduction from 2D rectangle parity to 2D range parity: On an insertion of a rectangle $R = [x_1, x_2] \times [y_1,y_2]$, we insert four points in the 2D range parity data structure, one roughly at each corner of $R$. The observation is that a point $q$ is inside $R$ iff it dominates only $x_1$ (if we ignore $q$ lying on one of the sides of $R$). Additionally, if $q$ is not inside $R$, then it dominates an even number of corners of $R$. Hence the parity of the number of rectangles stabbed is the same as the parity of the number of points dominated if we replace each rectangle with a point at each corner. The only minor issue is when $q$ lies on one of the sides of $R$. We give a full reduction also handling this case in the following:

Recall we require coordinates to be integer. From the rectangle $R$, we insert the four points $p_1=(2x_1, 2y_1), p_2=(2x_1,2y_2+1), p_3=(2x_2+1,2y_1), p_4=(2x_2+1,2y_2+1)$. For a 2D rectangle parity query point $q = (x,y)$, we ask the 2D parity counting query $(2x,2y)$. The factors of $2$ and the additive $+1$ is used to handle the ``on one of the sides'' case. The crucial observation is that 
$$
x \in [x_1, x_2] \Leftrightarrow 2x_1 \leq 2x \leq 2x_2 < 2x_2+1.
$$
Similarly 
$$
y \in [y_1,y_2] \Leftrightarrow 2y_1 \leq 2y \leq 2y_2 < 2y_2+1.
$$
Also, if $x>x_2$ then $2x > 2x_2+1$ and if $y>y_2$ then $2y>2y_2+1$ because all coordinates were integer.
It follows that if $q$ is inside the rectangle $R$, then $(2x,2y)$ dominates precisely $p_1$. If $x<x_1$ or $y<y_1$, then $(2x,2y)$ dominates none of the points $p_1,p_2,p_3$ and $p_4$. If $x>x_2$ but $y_1 \leq y \leq y_2$, then $(2x,2y)$ dominates precisely $p_1$ and $p_3$. If $x_1 \leq x \leq x_2$ and $y>y_2$, then $(2x,2y)$ dominates precisely $p_1$ and $p_2$. Finally, if $x>x_2$ and $y>y_2$, then $(2x,2y)$ dominates all the points $p_1,p_2,p_3$ and $p_4$. We conclude that $(2x,2y)$ dominates an odd number of points amongst $p_1,p_2,p_3$ and $p_4$ iff $q$ is inside the rectangle $R$. Hence the result of the 2D range parity query is the same as the result of the 2D rectangle parity query. This concludes the proof of Theorem~\ref{thm:reducerangeparity}.

\subsection{Range Selection and Range Median}
In this section, we show that a data structure for dynamic range selection solves parity searching in Butterfly graphs. Recall that in range selection, we are given an array $A=\{A[0],\dots,A[n-1]\}$ of integers, initially all $0$. A query is triple $(i,j,k)$. The goal is to return the index of the $k$'th smallest element in $\{A[i],\dots,A[j]\}$, breaking ties arbitrarily. We prove our lower bound for the special case of prefix selection in which we force $i=0$ and where we are required to return only whether the $k$'th smallest element is stored in an evenly indexed position or an odd one.

We start by describing our reduction in the setting where we need to return the whole index of the $k$'th smallest element, not just the parity. At the end, we argue that all we really need is the parity of the position. Our reduction starts by re-executing the reduction to 2D rectangle parity from Section~\ref{sec:rangeparity}. That is, each edge $(u,v)$ of a Butterfly graph $G_i$ is mapped to a rectangle $R^{(u,v)}_i$ such that a query $(s,t)$ to parity searching in Butterfly graphs must sum the weight $z^{(u,v)}$ of $(u,v)$ iff the point $(s,t)$ is inside the rectangle $R^{(u,v)}_i$. We will also use the fact that the proof of our lower bound for parity searching in Butterfly graphs is for a distribution where \emph{every} edge has its weight set before we query. Thus for our reduction, it suffices to show that we can solve the parity searching query $(s,t)$ after a sequence of updates that have set the weight of every edge of every $G_i$. Finally, we also exploit that the rectangles obtained by reduction from the Butterfly graphs have the property that $R^{(u,v)}_i$ and $R^{(x,y)}_j$ are disjoint if $i=j$ and the edges $(u,v)$ and $(x,y)$ are at the same depth of the Butterfly graph $G_i$. This implies that any point is contained in no more than $\Delta := \sum_{i=1}^\ell d_i = O(\lg^2 n)$ rectangles, one for each layer of each Butterfly ($d_i$ is the depth of Butterfly $G_i$). 

The instance we create is the concatenation of two arrays $B$ and $A$, i.e. $C := B \circ A$. By concatenation, we simply mean that $C[i] := B[i]$ in case $i \in [|B|]$ and otherwise $C[i] := A[i-|B|]$. 

\paragraph{The $B$ array.}
The array $B$ is quite simple. If $n$ denotes the total number of rectangles, then $B$ has $4n \Delta$ entries. We think of $B$ as being partitioned into batches of $(\Delta+1)$ entries, where the $j$'th batch consists of entries $B[(\Delta+1) j],B[(\Delta+1) j+1], \dots, B[(\Delta+1) (j+1) - 1]$. We set the entries of $B$ during epoch $\ell$ (the biggest epoch), such that each update of epoch $\ell$ sets $O(\Delta)$ entries of $B$. This means that the worst case update time of the data structure goes up by a factor $O(\Delta) = O(\lg^2 n)$. The values assigned to the entries of $B$ are as follows: In the $j$'th batch, we set the values to $B[(\Delta+1) j + i] \gets (\Delta+2) j + i+1$ for $i=0,\dots,\Delta$. 

\paragraph{The $A$ array.}
From each rectangle $R^{(u,v)}_i = [x_1, x_2] \times [y_1 , y_2]$, we define the four points $p_1=(2 x_1,2 y_1), p_2 = (2x_1, 2y_2+1), p_3=(2 x_2+1,2y_1)$ and $p_4 = (2x_2+1,2y_2+1)$. Then by the arguments of the previous section, a query point $(s,t)$ is inside $R^{(u,v)}_i$ iff $(2s,2t)$ dominates only $p_1$. Otherwise, $(2s,2t)$ dominates either precisely $\{p_1,p_2\}$, $\{p_1,p_3\}$ or $\{p_1,p_2,p_3,p_4\}$. Let $P$ be the collection of all points $p_1,p_2,p_3,p_4$ defined from the rectangles $R^{(u,v)}_i$ for every $i=1,\dots,\ell$ and every edge $(u,v) \in G_i$. Note that $P$ may contain duplicates. We keep the duplicates, so $P$ is multi-set and $|P|=4n$. Also observe that the collection $P$ is fixed and independent of any weights assigned to edges, it is merely a predefined set of points. For each point $p \in P$, define $\rank_x(p)$ to be the rank of $p.x$ amongst all $x$-coordinates of points in $P$. Since several points may have the same $x$-coordinate, we break ties in some arbitrary but fixed manner such that all ranks are unique integers from $0$ to $|P|-1$. Our maintained array $A$ has one entry for each point $p \in P$, namely entry $A[\rank_x(p)]$. We similarly define $\rank_y(p)$ for each $p$, such that $\rank_y(p)$ is the rank of $p.y$ amongst all $y$-coordinates of points in $P$. Again, we break ties in some arbitrary but fixed manner. Note that $\rank_x(p)$ and $\rank_y(p)$ are fixed and independent of any weights assigned to edges.

Initially all entries of $A$ are $0$. On an update setting the weight of $R^{(u,v)}_i$ to $1$, we do the updates:
\begin{itemize}
\item $A[\rank_x(p_1)] \gets \rank_y(p_1)(\Delta+2)$.
\item $A[\rank_x(p_2)] \gets |P|(\Delta+2)$.
\item $A[\rank_x(p_3)] \gets |P|(\Delta+2)$.
\item $A[\rank_x(p_4)] \gets \rank_y(p_4) (\Delta+2)$.
\end{itemize}
On an update setting the weight of $R^{(u,v)}_i$ to $0$, we instead do the updates:
\begin{itemize}
\item $A[\rank_x(p_1)] \gets |P|(\Delta+2)$.
\item $A[\rank_x(p_2)] \gets \rank_y(p_2) (\Delta+2)$.
\item $A[\rank_x(p_3)] \gets \rank_y(p_3) (\Delta+2)$.
\item $A[\rank_x(p_4)] \gets |P|(\Delta+2)$.
\end{itemize}
Conceptually, think of the above choices as inserting the points $p_1$ and $p_4$ on a weight of $1$, and inserting the points $p_2$ and $p_3$ on a weight of $0$. Setting the remaining entries to $|P|(\Delta+2)$ can be thought of as disregarding these entries in selection queries.

\paragraph{Answering the Query.}
On a query $(s,t)$, we compute the index $j \in [|P|]$ such that all points $p$ in $P$ with $p.x \leq 2s$ have $\rank_x(p) \leq j$ and all points $p$ in $P$ with $p.x > 2s$ have $\rank_x(p) > j$. Note that this index is completely determined from the query $(s,t)$ and does not depend on any weights assigned. Since computation is free in the cell probe model, computing $j$ is free of charge. Similarly, we can compute the index $h \in [|P|]$ such that all points $p \in P$ with $p.y \leq 2t$ have $\rank_y(p) \leq h$ and all points $p \in P$ with $p.y > 2t$ have $\rank_y(p)>h$. 

The crucial observation is that if we disregard the rectangles $R^{(u,v)}_i$ containing $(s,t)$ (i.e. disregard the corresponding entries in $A$), then we know exactly what value of $k$ that makes the query $(0,j+|B|,k)$ return the index $r:= h(\Delta+1)+\Delta$. Note that this index is the last index in the $h$'th batch of $B$. 

To see how one can determine this value of $k$ without knowing the weights, first note that the value of $k$ resulting in the answer $r$ to the query $(0,j+|B|,k)$, is precisely the value $k$ such that entry $B[r]$ stores the $k$'th smallest element amongst $B[0],\dots,B[|B|-1],A[0],\dots,A[j]$. We thus need to argue that the number of entries amongst $B[0],\dots,B[|B|-1],A[0],\dots,A[j]$ that stores a value less than or equal to $B[r] = h(\Delta+2)+\Delta + 1$ is independent of the weights assigned during updates (ignoring entries of $A$  corresponding to rectangles $R^{(u,v)}_i$ containing $(s,t)$ of course).

It is clear that the number of entries amongst $B[0],\dots,B[|B|-1]$ that stores a value less than or equal to $B[r]$ is independent of the weights since it is precisely the $r+1$ entries $B[0],\dots,B[r]$. Thus what remains is to argue that we can also compute the number of such entries amongst $A[0],\dots,A[j]$.

To compute it for $A[0],\dots,A[j]$, examine every rectangle $R^{(u,v)}_i$ not containing $(s,t)$. Let $p_1,p_2,p_3,p_4$ be the points defined from $R^{(u,v)}_i$ above. By the arguments earlier, we have that $(2s,2t)$ dominates either $\{p_1,p_2\}, \{p_1,p_3\}$ or $\{p_1,p_2,p_3,p_4\}$. Now observe that $\rank_x(p_i) \leq j$ iff $p_i.x \leq 2s$ and $\rank_y(p_i) \leq h$ iff $p_i.y \leq 2t$. This means that \emph{if} we perform the update $A[\rank_x(p_i)] \gets \rank_y(p_i)(\Delta+2)$, then $A[\rank_x(p_i)]$ is amongst entries $A[0],\dots,A[j]$ and stores a value less than or equal to $B[r] = h(\Delta+2)+\Delta+1$ iff $(2s,2t)$ dominates $p_i$. Also, if we instead performed the update $A[\rank_x(p_i)] \gets |P|(\Delta+2)$, then $A[\rank_x(p_i)] > B[r]$. Now examine the two different update strategies depending on whether the edge $(u,v)$ is assigned the weight $1$ or $0$. The crucial property of our reduction is that precisely one of $\{p_1,p_2\}$ is updated as $A[\rank_x(p_i)] \gets \rank_y(p_i)(\Delta+2)$. Similarly, precisely one of $\{p_1,p_3\}$ is updated as $A[\rank_x(p_i)] \gets \rank_y(p_i)(\Delta+2)$. Finally, precisely two of $\{p_1,p_2,p_3,p_4\}$ are updated as $A[\rank_x(p_i)] \gets \rank_y(p_i)(\Delta+2)$. Thus for each $R^{(u,v)}_i$ not containing $(s,t)$, we know exactly how many of the corresponding entries of $A$ that is amongst $A[0],\dots,A[j]$ and which store a value less than or equal to $B[r]$, independently of the weights assigned. Thus it follows that if we disregard entries of $A$ corresponding to rectangles containing $(s,t)$, we know the value $k$ such that the selection query $(0,j+|B|,k)$ returns the index $r = h(\Delta+1)+\Delta$.

Let $k$ be the value computed above (without any cell probes), i.e. $k$ is the integer such that the selection query $(0,j+|B|,k)$ returns the index $r=h(\Delta+1)+\Delta$. Recall that $B[r]$ is the last index in the $h$'th chunk of $B$, and this chunk consists of the entries $B[h(\Delta+1)+i] = h(\Delta+2)+i+1$ for $i=0,\dots,\Delta$. To answer the parity searching query $(s,t)$, we run the range selection query $(0,j+|B|,k)$. If the rectangles containing $(s,t)$ were not there, the data structure would return $r$. Now examine the change in the answer to the query $(0,j+|B|,k)$ as we perform the updates corresponding to rectangles $R^{(u,v)}_i$ containing $(s,t)$. For such rectangles, $(2s,2t)$ dominates exactly $\{p_1\}$ and thus by the arguments above, the four entries $A[\rank_x(p_1)],A[\rank_x(p_2)],A[\rank_x(p_3)],A[\rank_x(p_4)]$ corresponding to $p_1,p_2,p_3$ and $p_4$ have the property that precisely entry $A[\rank_x(p_1)]$ is amongst $A[0],\dots,A[j]$ and stores a value less than or equal to $B[r]$ if the weight assigned to $R^{(u,v)}_i$ is $1$. If the weight assigned to $R^{(u,v)}_i$ is $0$, then none of the entries have that property. Since we are selecting for the $k$'th smallest element in $B[0],\dots,B[|B|-1],A[0],\dots,A[j]$, we get that the first rectangle containing $(s,t)$ and being assigned the weight $1$ changes the answer to the query $(0,j+|B|,k)$ to $r-1$. The next changes the answer to $r-2$ and so forth. Crucially, we know that $(s,t)$ is contained in at most $\Delta$ rectangles. Therefore, the index returned will be exactly $r-m$ where $m$ is the number of rectangles containing $(s,t)$ and being assigned the weight $1$. The chunks of $\Delta+1$ entries thus prevents ``overflows'' in some sense, ensuring that each rectangle stabbed having weight $1$ decrements the returned index by exactly $1$. Since $r := h(\Delta+2)+\Delta + 1$ is completely determined from the query alone, and the data structure returns the value $r-m$, it follows that we can compute $((r-m) \mod 2) + (r \mod 2) = (-m \mod 2) = (m \mod 2)$, which is the answer to the parity searching query. Note that this computation needs \emph{only} the parity of the index $(r-m)$ returned by the query. This concludes the proof of Theorem~\ref{thm:reduceselect}.

\section*{Acknowledgement}
We are very grateful to Rocco Servedio and Oded Regev for insightful discussions on the Peak-to-Average Lemma, and in particular, 
to Alexander Sherstov for observing and sharing with us the proof of the lower bound (Claim \ref{cl_sherstov_hyperplane} in Appendix \ref{app_pf_tight_PtA}).

\bibliographystyle{alpha}
\bibliography{refs}

\appendix

\section{Proof of Lemma~\ref{lem:discrete_chebyshev}~\cite{BCWZ99}}\label{app_pf_cheby}
\ifx\mainfile\undefined
\documentclass{article}

\usepackage{fullpage, amsmath, amsthm, amssymb}
\usepackage{enumerate}
\usepackage{float}
\usepackage{color}

\newtheorem{definition}{Definition}
\newtheorem{theorem}{Theorem}
\newtheorem{corollary}{Corollary}
\newtheorem{lemma}{Lemma}
\newtheorem{proposition}{Proposition}
\newtheorem{claim}{Claim}
\newtheoremstyle{restate}{}{}{\itshape}{}{\bfseries}{~(restated).}{.5em}{\thmnote{#3}}
\theoremstyle{restate}
\newtheorem*{restate}{}

\newcommand{\algwidth}{0.97\textwidth}
\newcommand{\Fam}{\mathcal{F}}
\DeclareMathOperator{\mmod}{mod}
\renewcommand{\P}{\textbf{P}}
\newcommand{\Q}{\textbf{Q}}
\newcommand{\E}{\mathop{{}\mathbb{E}}}
\renewcommand{\S}{\textbf{S}}
\newcommand{\p}{\textbf{p}}
\renewcommand{\log}{\lg}

\newcommand{\R}{\mathbb{R}}
\newcommand{\eps}{\epsilon}
\newcommand{\path}{\mathsf{path}}
\newcommand{\suc}{\overrightarrow{\mathsf{adv}}}

\newcommand{\mnote}[1]{ \marginpar{\tiny\bf
            \begin{minipage}[t]{0.5in}
              \raggedright #1
           \end{minipage}}}
           
\newcommand{\GF}{GF}
\newcommand{\cA}{\mathcal{A}}
\newcommand{\cB}{\mathcal{B}}
\newcommand{\cC}{\mathcal{C}}
\newcommand{\cE}{\mathcal{E}}
\newcommand{\cI}{\mathcal{I}}
\newcommand{\cP}{\mathcal{P}}
\newcommand{\cG}{\mathcal{G}}
\newcommand{\cH}{\mathcal{H}}
\newcommand{\cM}{\mathcal{M}}
\newcommand{\cN}{\mathcal{N}}
\newcommand{\cX}{\mathcal{X}}
\newcommand{\cY}{\mathcal{Y}}
\newcommand{\cS}{\mathcal{S}}
\newcommand{\cQ}{\mathcal{Q}}
\newcommand{\cU}{\mathcal{U}}
\newcommand{\cT}{\mathcal{T}}
\newcommand{\cR}{\mathcal{R}}
\newcommand{\cL}{\mathcal{L}}
\newcommand{\cF}{\mathcal{F}}
\newcommand{\cZ}{\mathcal{Z}}
\newcommand{\cD}{\mathcal{D}}
\newcommand{\cJ}{\mathcal{J}}
\newcommand{\cK}{\mathcal{K}}
\newcommand{\cO}{\mathcal{O}}
\newcommand{\cW}{\mathcal{W}}

\newcommand{\bU}{\mathbf{U}}

\newcommand{\tS}{\tilde{S}}
\newcommand{\ORC}{\oplus\mathsf{ORC}}
\newcommand{\orc}{\mathsf{ORC}}
\newcommand{\poly}{\mathrm{poly}}
\newcommand{\ep}{\mathbf{U}}

\DeclareMathOperator{\supp}{supp}
\DeclareMathOperator{\rank}{rank}

\newcommand{\nz}{non-zero}
\newcommand{\AND}{\mathsf{AND}_k}

\begin{document}

\fi

\begin{restate}[Lemma~\ref{lem:discrete_chebyshev}]
For any $k$ and $M$ satisfying $2\leq M\leq 2^{O(k)}$, there exists a polynomial $Q=Q_{k,M}(x_1,\ldots,x_k)$ such that
\begin{enumerate}[(i)]
    \item
        $Q$ has total degree $O(\sqrt{k\log M})$;
    \item
        $|Q(0^k)|\geq M$;
    \item
        $\forall x\in\{0,1\}^k\setminus\{0^k\},\left|Q(x)\right|\leq 1$;
    \item
        The sum of absolute values of all coefficients is at most $\exp(\sqrt{k\log M})$.
\end{enumerate}
\end{restate}

The main idea is to design a lower degree univariate polynomial that takes large value at 0, takes small value at every integer between $1$ and $k$, and has small coefficients.
Then apply this polynomial on the Hamming weight of $x$, i.e., $x_1+\cdots+x_k$.
To design such a polynomial, we are going to use Chebyshev polynomials.\footnote{Note that the (standard) Chebyshev polynomials satisfying the second and third 
properties must have degree $\Theta(\sqrt{k}\log M)$, which is prohibitively large for our application.}
The high-level idea is to first take a Chebyshev polynomial $P$ that is large at $0$, and is small in $[t, k]$ for some parameter $t$.
Then for each integer $i$ between $1$ and $t-1$, we design another polynomial that vanishes at $i$, is not small at $0$, and is not large in $[i+1,k]$.
Finally, we multiply all these polynomials together, which will produce a polynomial with the claimed properties.

\begin{proof}

If $M\geq 2^{0.05k}$, the statement is trivial, as we can simply set $Q=M\cdot \prod_{i=1}^k (1-x_i)$.
Thus, in the following we are going to assume $M\leq 2^{0.05k}$.

In order to bound the sum of coefficients, let us define the following function on polynomials.
For a polynomial $P$ on $l$ variables such that $P(x)=\sum_{I\in\mathbb{N}^l} \alpha_Ix^I$, define $C(P, s)$ to be the sum of coefficients of $P$ weighted by exponentials of the total degrees:
\[
	C(P, s):= \sum_{I\in\mathbb{N}^l} |\alpha_I| s^{|I|}.
\]
Equivalently, $C(P, s)$ is $P$ evaluated on $x_i=s$ for all $i$, after replacing each coefficient by its absolute value.
Also, the last property in the statement is equivalent to $C(Q, 1)\leq \exp(\sqrt{k\log M})$.
It is easy to verify the function $C$ has the following properties:
\begin{enumerate}[(a)]
	\item
		$C(P_1+ P_2, s)\leq C(P_1, s)+ C(P_2, s)$;
	\item
		$C(P_1\cdot P_2, s)\leq C(P_1, s)\cdot C(P_2, s)$;
	\item
		For univariate $P_1$, $C(P_1\circ P_2, s)\leq C(P_1,C(P_2, s))$.
\end{enumerate}

Now consider the (univariate) Chebyshev polynomials $T_n$ recursively defined as follows:
\begin{itemize}
	\item
		$T_0(x)=1$;
	\item
		$T_1(x)=x$;
	\item
		$T_{n+1}(x)=2xT_{n}(x)-T_{n-1}(x)$.
\end{itemize}

It is well-known that $T_n$ is a degree-$n$ polynomial satisfying
\begin{itemize}
	\item
		$|T_n(x)|\leq 1$ for $-1\leq x\leq 1$;
	\item
		$|T_n(-1-\epsilon)|=|T_n(1+\eps)|\geq \frac{1}{2}e^{n\sqrt{\epsilon}}$ for $0\leq \epsilon<0.1$.
\end{itemize}
We have $C(T_0, s)=1$, $C(T_1, s)=s$, and 
\[
	C(T_{n+1}, s)\leq 2s\cdot C(T_n, s)+C(T_{n-1},s).
\]
By induction, we have $C(T_n, s)\leq (2s+1)^n$.
We are going to construct $Q$ based on Chebyshev polynomials.

Let $t$ be an integer parameter to be set later.
Consider the following (univariate) polynomials:
\begin{itemize}
	\item
		let $P(x)=T_n((x-k)/(k-t))$ for $n=\lceil\sqrt{k/t-1}\ln 2M\rceil$;
	\item
		For each $i=1,2,\ldots,t-1$, let $Q_i(x)=(T_n((x-k)/(k-i))-T_n(-1))/2$ for $n=\lceil2\sqrt{k/i-1}\rceil$.
\end{itemize}

It is easy to verify that, as long as $t\leq 0.05 k$, we have
\begin{itemize}
	\item
		$|P(0)|\geq M$, $|P(x)|\leq 1$ for $x\in [t, k]$;
	\item
		$C(P, k)\leq C(T_n,2k/(k-t))\leq 6^{\sqrt{k/t-1}\ln 2M+1}$ by Property $(c)$ above;
	\item
		$|Q_i(0)|\geq 1$, $Q_i(i)=0$ and $|Q_i(x)|\leq 1$ for $x\in [i+1,k]$;
	\item
		$C(Q_i,k)\leq (C(T_n,2k/(k-i))+1)/2\leq 6^{2\sqrt{k/i-1}+1}$.
\end{itemize}

Finally, we are going to set $t=\log M(\leq 0.05 k)$, and define the polynomial $Q$ as follows:
\[
	Q(x_1,\ldots,x_k):=(P\cdot Q_1\cdot Q_2\cdot\cdots\cdot Q_{t-1})(x_1+x_2+\cdots+x_k).
\]

In the following, we prove $Q$ has the claimed properties.
\begin{enumerate}[(i)]
	\item
		The degree of $Q$ is 
		\[
			O(\sqrt{k/t}\log M+\sum_{i=1}^{t-1}\sqrt{k/i})=O(\sqrt{k/t}\log M+\sqrt {kt})=O(\sqrt {k\log M}).
		\]
	\item
		$|Q(0^k)|=|P(0)\cdot Q_1(0)\cdots Q_{t-1}(0)|\geq M$.
	\item
		For $x$ such that $x_1+\cdots+x_k=|x|<t$, we have $Q(x)=0$, since $Q_{|x|}(x)=0$. 
		Otherwise, $|P(x)|\leq 1$ and $|Q_i(x)|\leq 1$.
		We also have $|Q(x)|\leq 1$.
	\item
		By Property $(b)$ above, we have $C(P\cdot Q_1\cdots Q_{t-1}, k)\leq \exp(\sqrt{k\log M})$.
		Thus, by Property $(c)$, we have 
		$$C(Q, 1)\leq C(P\cdot Q_1\cdots Q_{t-1},k)\leq \exp(\sqrt{k\log M})$$ as claimed.
\end{enumerate} 

\end{proof}

\ifx\mainfile\undefined

\bibliographystyle{alpha}
\bibliography{refs}
\end{document}
\fi


\section{Tightness of the Peak-to-Average Lemma} \label{app_pf_tight_PtA}
In this section we show that Lemma \ref{lem_peak_to_avg} is tight, in the sense that there is a function 
$f : \{-1,1\}^k \mapsto \R$ satisfying the premise of the lemma yet the average value of $f$ conditioned on any subset of 
$o(\sqrt{k\log 1/\eps})$ coordinates is $0$.
In other words, conditioning on $o(\sqrt{k\log 1/\eps})$ coordinates of $f$ provides \emph{no advantage at all} in predicting 
the value of $f$, despite the fact that $\|f\|_\infty \geq \eps$. 

The key for constructing our tight counter example  is the following claim, which relies on the well known fact that 
the $\eps$-approximate degree of the $\AND$ function is $\Theta(\sqrt{k\log 1/\eps})$.  
The following elegant corollary was pointed out to us by Alexander Sherstov (private communication): 

\begin{claim}\label{cl_sherstov_hyperplane}
For every $\eps>2^{-O(k)}$, there is a \emph{univariate} polynomial $Q: [k] \rightarrow \R$ of degree at most  
$k - \Omega(\sqrt{k\log 1/\eps})$ satisfying 
$$\left|Q(0)\right| > \eps\cdot \sum_{i=0}^k \left| {k\choose i}\cdot Q(i)\right| = \eps .$$  
\end{claim}

Let us first see why this claim implies the existence of our desired function $f$. To this end, assume w.l.o.g that $k$ is even and 
let $Q$ be the univariate polynomial obtained from Claim \ref{cl_sherstov_hyperplane}, and define the \emph{multivariate} polynomial 
$f_Q:\{-1,1\}^k\rightarrow \R$ as \[ f_Q(x_1,\ldots,x_k) := Q(|x|)\cdot \prod_{i=1}^k x_i , \]
where $|x|:= k/2 + \sum_{i=1}^k  x_i/2 $ is the number of $1$ coordinates in $x\in \{-1,1\}^k$. By Claim \ref{cl_sherstov_hyperplane}, 
the total degree of $Q$ (now as a \emph{multivariate} polynomial) is still at most $r := k - \Omega(\sqrt{k\log 1/\eps})$. 
Writing $Q(|x|)$ as the sum of its monomials and recalling that $x_i^2=1$ over $\{-1,1\}$, observe that multiplying 
each monomial by $\prod_{i=1}^k x_i$ simply ``inverts" each monomial $\prod_{i\in S}x_i$ to the monomial $\prod_{i\notin S}x_i$. 
Since each monomial of $Q$ was of degree at most $r = k - \Omega(\sqrt{k\log 1/\eps})$, this implies that all monomials 
of $f_Q$ are of degree \emph{at least} $k - r = \Omega(\sqrt{k\log 1/\eps})$. In particular, it follows that for any 
subset of coordinates $Y\subset [k]$, $|Y| = o\left(\sqrt{k\cdot \log 1/\eps}\right)$, the average value of $f_Q$ conditioned on coordinates in $Y$ is 
\[ \sum_{y\in\{-1,1\}^Y}\left|\sum_{x|_Y=y}f_Q(x)\right| = 0 ,\]
since each monomial of $f_Q$ is of degree $\Omega(\sqrt{k\log 1/\eps})$, hence conditioning on any 
$o\left(\sqrt{k\cdot \log 1/\eps}\right)$ of its coordinates has $0$ correlation with the value of the monomial. 

To see why $f_Q$ satisfies the two premises of Lemma \ref{lem_peak_to_avg}, first observe that the conclusion of Claim \ref{cl_sherstov_hyperplane} 
implies that $\|f_Q\|_1 = \sum_{i=0}^{k}  {k \choose i}\cdot |Q(i)| = 1$, as required by the first condition of the lemma. To see why the second premise of 
the lemma holds, we have 
\[  |f_Q(-1,\ldots , -1)| :=  |Q(0)|\cdot \left|\prod_{i=1}^k x_i \right| = |Q(0)|  >  \eps \; , \]
where the last inequality follows from Claim \ref{cl_sherstov_hyperplane}.   
In particular, $\max_{x}|f_Q(x)| \geq |f_Q(-1,\ldots , -1)|\geq \eps$, as the premise of the Peak-to-Average Lemma requires. 
Hence, to complete the proof of our lower bound, it remains to prove Claim \ref{cl_sherstov_hyperplane}. 

\begin{proof}[Proof of Claim \ref{cl_sherstov_hyperplane}]
Thought the proof we represent polynomials $P: [k] \mapsto \R$ by their corresponding ``truth tables" over the domain $[k]$ (i.e., a vector in $\R^{k+1}$). 
For example, the function $\AND : [k] \mapsto \{0,1\}$ is represented by the vector $(1,0,\ldots , 0)$. 
Denote by $\cB_{\eps}(\AND)$ the $\ell_\infty$ ball of radius $\eps$ centered at $(1,0,\ldots , 0)$, i.e., the convex set of (truth-tables of) functions that  
point-wise $\eps$-approximate the $\AND$ function on $[k]$, and  
denote by $\cP_{d}$ the set of all (truth-tables of) real polynomials of degree $\leq d$ over the domain $[k]$. It is not hard to 
verify that under our representation, the set $\cP_d$ is also convex (since degree-$d$ polynomials are closed under convex 
combinations). It is well known (e.g., Theorem 3 in \cite{BCWZ99}) that 
the lowest degree of a \emph{univariate real} polynomial that point-wise approximates the $\mathsf{AND}_k$ function to within additive error $\eps$ 
is $\Theta(\sqrt{k\log 1/\eps})$  (for every $\eps>2^{-O(k)}$), 
hence if we set $d =\frac{1}{C}\cdot \sqrt{k\log 1/\eps}$ for a large enough constant $C$,  the two (convex) sets $\cB_\eps(\AND)$ 
and $\cP_d$ are \emph{disjoint}. In this case, Farkas' Lemma implies that there is some \emph{separating hyperplane} $\psi \in \R^{k+1}$ of unit norm 
($\|\psi\|_1=1$) and some $b\geq 0$, for which: 
$(i) \; \langle  P,\psi \rangle \leq b$ for all $P\in \cP_d$; and $(ii) \; \langle  g,\psi \rangle > b \geq 0$ for all $g\in \cB_\eps(\AND)$.
Note that the first condition actually implies that $\psi$ must be \emph{orthogonal} to all of $\cP_d$, i.e., $\langle  P,\psi \rangle = 0$, 
since degree-$d$ polynomials are closed under scaling and negations (so if $\langle  P,\psi \rangle = b \neq 0$ for some degree-$d$ polynomial $P$, 
then $\langle  c\cdot P,\psi \rangle = c\cdot b$ for any $c>0$,  
and $c\cdot P\in \cP_d$). Moreover, the second condition in particular holds for the vector $g^* \in \cB_\eps(\AND)$ defined by $g^*(i):=-sgn(\psi(i))\cdot \eps$ 
for every $i\neq 0$  and $g^*(0) :=1-\eps$, in which case $\langle  g^*,\psi \rangle = (1-\eps)\psi(0) - \eps\cdot\sum_{i=0}^k \psi(i) =  |\psi(0)|- \eps\cdot\|\psi\|_1$. 
In conclusion, from the above two conditions we can derive that: 
\begin{enumerate}
\item[(1)] $\langle  P,\psi \rangle = 0$ for all $P\in \cP_d$; 
\item[(2)] $  |\psi(0)| > \eps\cdot \|\psi\|_1$ ;
\item[(3)] $\|\psi\|_1=1$ .
\end{enumerate} 
It is a well known fact\footnote{This is a straightforward corollary of the combinatorial 
identity $\sum_i {k\choose i} (-1)^i (a_0 + a_1i+a_2i^2\ldots + a_k i^k) = (-1)^k k! a_k$ (for arbitrary $(a_0,\ldots , a_k)$, see e.g., Equation (5.42) in \cite{ConcreteMath94}). 
Indeed, for any $P\in \cP_d$ and $\psi$ of the form \eqref{eq_orth_poly_form}, $\sum_i P(i)\cdot \psi(i) = {k\choose i} (-1)^i (Q\cdot P)(i)$, so whenever $deg(Q) < k-d $, 
we have $deg(P\cdot Q) < k$ hence the $k$th coefficient is $0$ and the above identity implies $\langle P,\psi \rangle= 0$ 
for any $P\in P_d$. The other direction follows from the observation that 
the orthogonal subspace to $\cP_d$ has dimension $k-d$, and indeed the space spanned by the polynomials in \eqref{eq_orth_poly_form} (ranging over 
all $Q$'s of degree at most $k-d$) has the latter dimension.} 
that every $\psi$ that satisfies (1) must be of the form: 
\begin{equation}\label{eq_orth_poly_form}
\psi(i) = {k\choose i} (-1)^i \cdot Q(i), 
\end{equation}
for all $i\in [k]$, where $Q$ is some \emph{degree $\leq (k-d)$-polynomial} (again, recall we are representing polynomials by their corresponding truth tables). 
Substituting this fact in (2), we conclude that there is a polynomial of degree at most $k-d = k - \Omega(\sqrt{k\log 1/\eps})$, satisfying 
\begin{align*}
&\left|{k\choose 0} (-1)^0 \cdot Q(0)\right| > \eps\cdot \sum_{i=0}^k \left| {k\choose i} (-1)^i \cdot Q(i)\right|. \\
\Leftrightarrow & \left|Q(0)\right| > \eps\cdot \sum_{i=0}^k \left| {k\choose i}\cdot Q(i)\right| = \eps , 
\end{align*}
since $\sum_{i=0}^k \left| {k\choose i}\cdot Q(i)\right| = \|\psi\|_1 = 1$, as claimed. 
\end{proof}


\end{document}